\documentclass[useAMS,usenatbib]{mn2e}
\usepackage{graphicx,amssymb,times}

\title[3-mm spectral imaging of the CMZ]{Spectral imaging of the Central 
Molecular Zone in multiple 3-mm molecular lines}
\author[P. A. Jones et al.]{P. A. Jones$^{1,2}$
\thanks{E-mail:pjones@phys.unsw.edu.au},
M. G. Burton$^{1}$, M. R. Cunningham$^{1}$, M. A. Requena-Torres$^{3}$, 
\newauthor{K. M. Menten$^{3}$, P. Schilke$^{4}$,
A. Belloche$^{3}$, S. Leurini$^{3}$, J. Mart\'in-Pintado$^{5}$, J. Ott$^{6}$,}
\newauthor{A. J. Walsh$^{7}$ }
\\
$^{1}$School of Physics, University of New South Wales, NSW 2052, Australia \\ 
$^{2}$Departamento de Astronom\'{\i}a, Universidad de Chile, Casilla 36-D, 
Santiago, Chile \\
$^{3}$Max-Planck-Institut f\"{u}r Radioastronomie, Auf dem H\"{u}gel 69, 53121 
Bonn, Germany \\
$^{4}$I. Physikalisches Institut, Universit\"{a}t zu K\"{o}ln, Z\"{u}lpicher Str. 77, 
50937 Köln, Germany \\
$^{5}$Centro de Astrobiolog\'{\i}a (CSIC/INTA), Ctra. de Torrej\'on a Ajalvir km 4, 
28850 Torrej\'on de Ardoz, Madrid, Spain \\
$^{6}$National Radio Astronomy Observatory, P.O. Box O, 1003 Lopezville Road, 
Socorro, NM 87801, USA \\
$^{7}$Centre for Astronomy, School of Engineering and Physical Sciences, James
Cook University, QLD, 4814, Australia \\
}

\begin{document}

\date{Accepted 2011 October 2. Received 2011 September 29; in original form 
2011 July 11}

\pagerange{\pageref{firstpage}--\pageref{lastpage}} \pubyear{2011}

\maketitle

\label{firstpage}

\begin{abstract}

We have mapped 20 molecular lines in the Central Molecular Zone
  (CMZ) around the Galactic Centre, emitting from 85.3 to 93.3~GHz.  This
  work used the 22-m Mopra radio telescope in Australia, equipped with
  the 8-GHz bandwidth UNSW--MOPS digital filter bank, obtaining $\sim 2$ 
  km~s$^{-1}$ spectral and $\sim 40$~arcsec spatial resolution.  The
  lines measured include emission from the c-C$_{3}$H$_{2}$,
  CH$_{3}$CCH, HOCO$^{+}$, SO, H$^{13}$CN, H$^{13}$CO$^{+}$, SO,
  H$^{13}$NC, C$_{2}$H, HNCO, HCN, HCO$^{+}$, HNC, HC$_{3}$N,
  $^{13}$CS and N$_{2}$H$^{+}$ molecules. The area covered is Galactic
  longitude $-0.7$ to $1.8$~deg. and latitude $-0.3$ to $0.2$~deg.,
  including the bright dust cores around Sgr~A, Sgr~B2, Sgr~C and
  G1.6-0.025.  We present images from this study and conduct a
  principal component analysis on the integrated emission from the
  brightest 8 lines. This is dominated by the first component, showing
  that the large-scale distribution of all molecules are very similar.
  We examine the line ratios and optical depths in selected apertures
  around the bright dust cores, as well as for the complete mapped
  region of the CMZ.  We highlight the behaviour of the bright HCN,
  HNC and HCO$^{+}$ line emission, together with that from the $^{13}$C
  isotopologues of these species, and compare the behaviour with that
  found in extra-galactic sources where the emission is unresolved
  spatially.  We also find that the isotopologue line ratios (e.g.
  HCO$^{+}$/H$^{13}$CO$^{+}$) rise significantly with increasing
  red-shifted velocity in some locations.
Line luminosities are also calculated and compared
to that of CO, as well as to line luminosities determined for external galaxies.
\end{abstract}


\section{Introduction}
\label{sec:intro}


The Central Molecular Zone (CMZ) is the region within about 300 parsec 
(2 degrees) of the Galactic Centre \citep{mose96} with a strong
concentration of molecular gas. The molecular distribution is most clearly
traced by CO \citep[e.g.][]{ba+87,ba+88,bi+97,ok+98,da+97,ok+07,ma+04}, which shows 
high gas densities, large shear velocities in the Galactic Centre potential well
with non-circular motions, large internal velocity dispersions within the clouds 
and high gas temperatures. The molecular mass in the CMZ is estimated as 
$(3^{+2}_{-1}) \times 10^{7}$~M$_{\sun}$ by \citet{da+98} from a range of 
tracers.

Some of the most prominent features \citep{pago96} of the CMZ are 
Sagittarius A (Sgr~A) the Galactic Centre cloud, 
Sagittarius B2 (Sgr~B2) and Sagittarius C (Sgr~C), named from the radio and near
infrared peaks \citep{pimi51, le62, hofreem71}.

The CMZ is asymmetric about the Galactic Centre, with the molecular distribution
biased towards positive longitudes, with Sgr~B2 a major feature at 
$l = +0.68^{\circ}$. The three-dimensional shape has been modelled as an 
elongated structure
e.g. by \citet{sa+04} using constraints from the CO emission and OH absorption,
with the positive longitude end closer to us.

The CMZ is prominent in far-infrared and sub-millimetre dust emission,
e.g. JCMT SCUBA observations at 850 and 450 $\mu$m \citep{pi+00}, APEX
ATLASGAL at 870 $\mu$m \citep{sc+09}, Bolocam at 1100 and 350 $\mu$m
\citep{ba+10} and HiGAL\footnote{https://hi-gal.ifsi-roma.inaf.it/higal/} 
at 70, 170, 250, 350 and 500 $\mu$m. 
This traces the $T \sim$20~K dust associated with the molecular 
gas with \citet{pi+00} estimating the total gas mass in the CMZ as 
$(5.3 \pm 1.0) \times 10^{7}$~M$_{\sun}$ from the dust emission.

The CMZ is a site of recent star formation \citep{yu+10} and very high
stellar densities. Because of the very high extinction and crowding, the 
stellar objects are best studied in the infrared and with high resolution,
e.g. see Spitzer studies at 24~$\mu$m \citep{hi+09} and 3.6 to 8.0~$\mu$m 
\citep{ra+08}.

The molecular medium of the CMZ has a surprisingly rich chemistry. The Sgr~B2 
region is well-studied \citep[e.g.][and references therein]{jo+08,jo+11} 
and known to have rich organic 
chemistry, particularly the Sgr~B2 (N) and (M) hot cores \citep{heva09}.
However, complex molecules are extended over the larger CMZ area as shown
by widespread CH$_{3}$OH (methanol) at 834~MHz detected by \citet{go+79},
HNCO (isocyanic acid) at 110~GHz imaged by \citet{da+97} and complex organic
molecules detected in individual clouds by \citet{re+08}. The chemistry in 
the CMZ
is similar to that in hot molecular cores (HMCs, hot star-forming dense cores
within giant molecular cloud complexes) but over a much larger scale 
\citep{re+06}. 

Other molecules imaged in the CMZ include SiO \citep{ma+97}, which traces 
extensive shocks, CS \citep*{tshauk99}, 
NH$_{3}$ \citep{na+09}, HCN \citep{ja+96}, HCO$^{+}$ and H$^{13}$CO$^{+}$
\citep{ri+10a}
and OH \citep{boco94} in absorption against the continuum.

It was argued by \citet{hu98} that ``large scale surveys in as many molecules, 
isotopomers and transitions as possible are essential to understand the 
structure of the CMZ'' preferably with resolution $< 2$~pc ($< 50$~arcsec).
The reason is that the distribution of a particular line depends on the
excitation, optical depth and abundance. The $^{12}$CO 1~--~0 line, in 
particular, does not trace the density structure well. 
To obtain the density and temperature distribution of the gas, multilevel 
studies, 
isotope studies and surveys in rarer, weaker molecular species are required.

We aim, in part, to fill this need.
We present here observations of the CMZ of 20 lines in the 3-mm band (including
two hydrogen recombination lines),
with resolution around 40 arcsec. Some preliminary results from this project
have been presented in \citet*{jobucu08} and \citet*{jobulo08}.

This is part of a larger project in which we are also observing the CMZ area
in multiple lines in the 7-mm band, with resolution around 65 arcsec.
Additionally we have imaged the CMZ area in multiple lines in the 12-mm band,
with resolution around 120 arcsec, as part of the wider H$_{2}$O southern 
Galactic Plane Survey (HOPS) \citep{wa+08,wa+11}. 

Due to the particular chemical complexity, and line strength, of the Sgr~B2
region, we have multi-line spectral images of this smaller sub-area of the CMZ
over the full 3-mm band from 82 to 114~GHz,
and 7-mm band from 30 to 50~GHz, presented in \citet{jo+08} and \citet{jo+11}
respectively.

We use the distance to the Galactic Centre of 8.0~kpc \citep{re+09}, so 
1~arcmin corresponds to 2.3~pc.

\section[]{Observations and Data Reduction}
\label{sec:obs}

The observations were made with the 22-m Mopra radio telescope, of the 
Australia Telescope National Facility (ATNF) using the MOPS digital filterbank.
The Mopra MMIC (Monolithic Microwave Integrated Circuit) receiver has a 
bandwidth of 8~GHz, and the MOPS backend can
cover the full 8-GHz range simultaneously in the broad band mode. This gives 
four 2.2-GHz sub-bands each with 8192 channels of 0.27~MHz. 
The lines in the CMZ are broad, so that the 0.27~MHz 
channels, corresponding to around 0.9~km~s$^{-1}$, are quite adequate. 

The Mopra receiver covers the range 77 to 117~GHz in the 3-mm band. We chose
the tuning centred at 89.3~GHz, to cover the range 85.3 to 93.3~GHz, giving the
spectral lines summarised in Table \ref{tab:lines_table}. This range was chosen
to include the strong lines of HCN, HCO$^{+}$ and HNC, which are some of 
the strongest lines in the 3-mm band (after $^{12}$CO and $^{13}$CO), plus
a good range of other lines.


The area was observed with on-the-fly (OTF) mapping, in blocks of 
$5.1 \times 5.1$ 
arcmin$^{2}$, in a similar way to that described in \citet{jo+08}.
We used position switching for bandpass calibration with
the off-source reference position observed before each 5 arcmin long source
scan. The reference position
(($\alpha, \delta)_{\rm J2000} = 17^{\rm h}51^{\rm m}03\rlap.$\,$^{s}$\,$6, 
-28^{\circ}22'47''$, or
$l = 1.093$ deg., $b = -0.735$ deg.), was carefully selected. It is hard
to balance the compromise of positions which are relatively emission free, in 
the Galactic Centre area, without too large a position offset for good spectral 
baselines.
The OTF maps used scan rate 4 arcsec/second, with 12 arcsec line spacing,
taking a little over an hour per block. We made pointing observations
of SiO maser positions (AH Sco or VX Sgr), before every map, to correct the 
pointing to within 
about 10 arcsec accuracy.  The system temperature was calibrated 
with a continuous noise diode, and ambient-temperature load (paddle) about 
every 30 minutes. The system temperatures were mean 210 to 225~K across the
85 to 93~GHz band, with standard deviation 50~K, depending on elevation
and weather conditions, and extremes 155 to 410~K.

The $5 \times 5$ arcmin$^{2}$ blocks were observed twice each with Galactic
latitude and longitude scan directions, slightly offset, 
to reduce scanning direction stripes, and to 
improve the signal to noise. The overall $2.5 \times 0.5$ deg$^{2}$ area was 
covered with a $30 \times 6$ grid of blocks, separated by 5 arcmin steps, 
making 360 OTF observations
required. In practice close to 400 OTF observations were used, including some
areas with OTF maps stopped by weather or other problems, and re-observed.
The area covered is between -0.72 to 1.80 degrees Galactic longitude, and 
-0.30 to 0.22 degrees Galactic latitude.

The observations were spread over three southern winter seasons, in 2007, 2008 
and 2009. Although the blocks worst affected by poor weather were re-observed,
the range of conditions during the different observations means that there
are inevitably some blocks with greater noise level than others.

The OTF data were turned into FITS data cubes with the {\sc livedata} and
{\sc gridzilla} 
packages\footnote{http://www.atnf.csiro.au/people/mcalabre/livedata.html}. 
The raw spectra in 
RPFITS\footnote{http://www.atnf.csiro.au/computing/software/rpfits.html} 
format,
were bandpass corrected and calibrated using the off-source reference spectra 
with {\sc livedata}, then
a robust second order polynomial fitted to the baseline and subtracted, with
the data output formatted
as SDFITS \citep{ga00} spectra. These spectra were then gridded into 
datacubes using
{\sc gridzilla}, with a median filter for the interpolation,
and combination of data over-sampled in position. The median was 
used, as this is more robust to the outliers caused by bad data. The scripts 
used for gridding allowed the lines (Table \ref{tab:lines_table}) to be specfied,
with their rest frequencies, so the {\sc gridzilla} output was FITS cubes with
velocity coordinates, combining data for the whole mapped area.

The FITS cubes were then read into the {\sc miriad} package for further
processing and analysis. In particular, as the emission is typically
of low surface brightness, the data were smoothed in velocity, with a 7-point
hanning kernel, to make a version of the data cubes with improved surface
brightness sensitivity\footnote{The sensitivity of brightness temperature in K
is improved by increasing the channel bandwidth.}. 
This gives around 1.07~MHz, or 3.6~km~s$^{-1}$ effective 
spectral resolution, which we use as a reduced size data cube by dropping every
second original pixel to make a Nyquist sampled version\footnote{The 7-point 
hanning smoothing gives FWHM four pixels, so dropping alternate pixels gives 
two pixels per FWHM.} with 0.54~MHz or 
1.8~km~s$^{-1}$ pixels. This is still quite adequate spectral resoution for the
broad lines in the CMZ, which are $> 10$~km~s$^{-1}$ wide, 
but for the rare narrow spectral feature, we use the
data cubes in the original 0.27~MHz or 0.9~km~s$^{-1}$ pixel versions.

The resolution of the Mopra beam varies between 36 arcsec at 86~GHz and
33 arcsec at 115~GHz \citep{la+05}, so the resolution in the final data
is around 39 arcsec, after the effect of the median filter convolution 
in the gridzilla interpolation. The main beam efficiency of Mopra 
varies between 0.49 at 86~GHz and 0.44 at 100~GHz
\citep{la+05}. 
However, the Mopra beam has substantial sidelobes \citep{la+05} so that the
extended beam efficiency, appropriate for the extended emission we are 
mostly studying here, is quite different (0.65 at 86~GHz).
Since we are largely concerned in this paper with the spatial
and velocity structure, we have mostly left the intensities 
in the T$_{A}^*$ scale,
without correction for the beam efficiency onto the T$_{MB}$ scale, except
for some of the quantitative analysis sections.

\begin{table}
\begin{center}
\caption{The lines imaged here in the 85.3 to 93.3~GHz range. The multiple
components listed here are, in general, blended due the large velocity
widths in the CMZ area. The observations covered the whole  85.3 to 93.3~GHz 
range, so do include many more weaker lines, particularly in Sgr B2,
as listed in \citet{jo+08}.}
\label{tab:lines_table}
\begin{tabular}{cccc}
\hline
Rough     & line ID          &                    & Exact      \\
Freq.     & molecule         & transition         & Rest Freq. \\
GHz       & or atom          &                    & GHz        \\
\hline
 85.34 &  c-C$_{3}$H$_{2}$   & 2(1,2) -- 1(0,1)    &  85.338906   \\
 85.46 &  CH$_{3}$CCH        & 5(3) -- 4(3)          &  85.442600   \\
       &                     & 5(2) -- 4(2)          &  85.450765   \\
       &                     & 5(1) -- 4(1)          &  85.455665   \\
       &                     & 5(0) -- 4(0)          &  85.457299   \\
 85.53 &  HOCO$^{+}$         & 4(0,4) -- 3(0,3)    &  85.531480   \\
 85.69 &  H                  & RRL H 42 $\alpha$   &  85.68839    \\
 86.09 &  SO                 & 2(2) -- 1(1)        &  86.093983   \\
 86.34 &  H$^{13}$CN         & 1~--~0 F=1-1        &  86.338735   \\
       &                     & 1~--~0 F=2-1        &  86.340167   \\
       &                     & 1~--~0 F=0-1        &  86.342256   \\
 86.75 &  H$^{13}$CO$^{+}$   & 1~--~0              &  86.754330   \\
 86.85 &  SiO                & 2~--~1 v=0          &  86.847010   \\
 87.09 &  HN$^{13}$C         & 1~--~0 F=0-1        &  87.090735   \\
       &                     & 1~--~0 F=2-1        &  87.090859   \\
       &                     & 1~--~0 F=1-1        &  87.090942   \\
 87.32 &  C$_{2}$H           & 1~--~0 3/2-1/2 F=2-1 & 87.316925   \\
       &                     & 1~--~0 3/2-1/2 F=1-0 & 87.328624   \\
 87.40 &  C$_{2}$H           & 1~--~0 1/2-1/2 F=1-1 & 87.402004   \\
       &                     & 1~--~0 1/2-1/2 F=0-1 & 87.407165   \\
 87.93 &  HNCO               & 4(0,4) -- 3(0,3)    &  87.925238   \\
 88.63 &  HCN                & 1~--~0 F=1-1        &  88.6304157  \\
       &                     & 1~--~0 F=2-1        &  88.6318473  \\
       &                     & 1~--~0 F=0-1        &  88.6339360  \\
 89.19 &  HCO$^{+}$          & 1~--~0              &  89.188526   \\
 90.66 &  HNC                & 1~--~0 F=0-1        &  90.663450   \\
       &                     & 1~--~0 F=2-1        &  90.663574   \\
       &                     & 1~--~0 F=1-1        &  90.663656   \\  	
 90.98 &  HC$_{3}$N          & 10~--~9             &  90.978989   \\
 91.99 &  CH$_{3}$CN         & 5(3)~--~4(3) F=6-5  &  91.971310   \\
       &                     & 5(3)~--~4(3) F=4-3  &  91.971465   \\
       &                     & 5(2)~--~4(2) F=6-5  &  91.980089   \\
       &                     & 5(1)~--~4(1)        &  91.985316   \\
       &                     & 5(0)~--~4(0)        &  91.987089   \\
 92.03 &  H                  & RRL H 41 $\alpha$   &  92.034434   \\
 92.49 &  $^{13}$CS          & 2~--~1              &  92.494303   \\
 93.17 &  N$_{2}$H$^{+}$     & 1~--~0 F$_1$=1-1 F=0-1   &  93.171621   \\
       &                     & 1~--~0 F$_1$=1-1 F=2-2   &  93.171917   \\
       &                     & 1~--~0 F$_1$=1-1 F=1-0   &  93.172053   \\
       &                     & 1~--~0 F$_1$=2-1 F=2-1   &  93.173480   \\
       &                     & 1~--~0 F$_1$=2-1 F=3-2   &  93.173777   \\
       &                     & 1~--~0 F$_1$=2-1 F=1-1   &  93.173967   \\
       &                     & 1~--~0 F$_1$=0-1 F=1-2   &  93.176265   \\
\hline
\end{tabular}
\end{center}
\end{table}

\section[]{Results}
\label{sec:results}

\subsection{Line data cubes}
\label{subsec:cubes}

We have data cubes for the twenty\footnote{As the observations included the 
full 8-GHz range, there are other weaker lines detected, mostly in Sgr~B2, but 
as they are quite weak and not extended over the larger area of the CMZ, we do 
not consider them further here. See Table 3 of \citet{jo+08} for a listing
of these Sgr~B2 lines.}
strongest lines in the 85.3 to 93.3~GHz range,
as listed in Table \ref{tab:lines_table}. The line identifications with 
rest frequencies and transitions are taken from the online NIST 
catalogue\footnote{http://physics.nist.gov/PhysRefData/Micro/Html/contents.html}
of lines known in the interstellar medium \citep{lodr04} and the splatalogue 
compilation\footnote{http://www.splatalogue.net/}.

The root-mean-square (RMS) noise level
and peak brightness are listed in Table \ref{tab:summary_table}. The RMS noise
is around 50 mK, except for the two lines of C$_{2}$H at 87.32 and 87.40~GHz
which are close to the edge of the sub-band at 87.3~GHz, where there is poor 
data. 

\begin{figure}
\includegraphics[angle=-90,width=8.0cm]{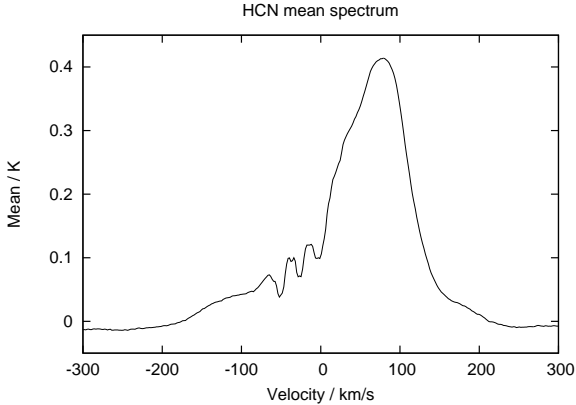}
\caption{Spectrum of HCN 1-0 averaged over the whole $2.5 \times 0.5$~deg$^2$ 
CMZ area observed.}
\label{fig:HCN_spec}
\end{figure}

The velocity range listed in Table \ref{tab:summary_table} indicates the range
of significant emission detected in the data cubes, and was the range over 
which we integrated to get images of the total line emission. 
We show the spectrum of HCN averaged over the whole CMZ area in Fig. 
\ref{fig:HCN_spec}, which has strong emission, seen over the largest velocity 
range. The range of velocities is large, due to the large line widths in
the CMZ and the large velocity gradient across the CMZ in the deep potential
well of the Galactic Centre. This makes the integrated emission image quite
sensitive to low level baselevel offsets, which while small in terms of 
brightness (e.g. of order 10 mK) becomes significant (e.g. of order 
K~km~s$^{-1}$) when integrated over the velocity range (order 100~km~s$^{-1}$).
See section \ref{subsec:integ_em} for discussion of the integrated emission.

\begin{table*}
\begin{center}
\caption{Statistics of the data cubes. The RMS noise values are  
from the hanning smoothed data with 1.07~MHz pixels: the original
0.27~MHz pixel data have noise a factor of 2 greater. Both the RMS noise and 
peak brightness temperature are in $T_{A}^{*}$. The velocity range
is that with line emission significantly above the noise level, and is the 
range used to integrate to get total line emission images. We also quote the 
position and velocity of the peak pixel.}
\label{tab:summary_table}
\begin{tabular}{cccccccc}
\hline
Line      & Molecule         & RMS       & Velocity & Peak & \multicolumn{2}{c}{Peak} & Peak \\
Freq.     & or atom ID       & noise     & range  &        & lat. & long. & vel. \\
GHz       &                  & mK        & km~s$^{-1}$ & K & deg. & deg.  & km s$^{-1}$ \\
\hline
 85.34 &  c-C$_{3}$H$_{2}$   &  54  & ~$-$80, 140 &  0.74 & 359.874 & -0.081 &  13 \\
 85.46 &  CH$_{3}$CCH        &  48  & ~$-$80, 140 &  0.75 & ~~0.663 & -0.036 &  65 \\
 85.53 &  HOCO$^{+}$         &  48  & $-$100, 120 &  0.89 & ~~0.684 & -0.013 &  69 \\
 85.69 &  H 42 $\alpha$      &  44  & $-$100, 100 &  0.40 & ~~0.668 & -0.036 &  70 \\
 86.09 &  SO                 &  49  & ~$-$10, ~90 &  1.32 & ~~0.668 & -0.035 &  58 \\
 86.34 &  H$^{13}$CN         &  54  & $-$160, 200 &  1.02 & 359.987 & -0.074 &  44 \\
 86.75 &  H$^{13}$CO$^{+}$   &  54  & $-$120, 180 &  0.90 & ~~0.663 & -0.035 &  52 \\
 86.85 &  SiO                &  58  & $-$100, 200 &  1.42 & 359.883 & -0.078 &  13 \\
 87.09 &  HN$^{13}$C         &  56  & ~$-$50, 100 &  0.72 & ~~0.665 & -0.029 &  53 \\
 87.32 &  C$_{2}$H           &  83  & $-$160, 80* &  1.45 & ~~0.674 & -0.022 &  76 \\
 87.40 &  C$_{2}$H           &  80  & $-$180, 120 &  0.73 & 359.875 & -0.086 &  14 \\
 87.93 &  HNCO               &  56  & $-$140, 170 &  5.07 & ~~0.692 & -0.022 &  71 \\
 88.63 &  HCN                &  42  & $-$220, 220 &  4.74 & ~~0.108 & -0.084 &  61 \\
 89.19 &  HCO$^{+}$          &  43  & $-$200, 220 &  3.79 & 359.618 & -0.246 &  19 \\
 90.66 &  HNC                &  46  & $-$200, 200 &  3.36 & 359.873 & -0.073 &  52 \\
 90.98 &  HC$_{3}$N          &  50  & $-$130, 180 &  3.90 & ~~0.663 & -0.033 &  57 \\
 91.99 &  CH$_{3}$CN         &  45  & $-$120, 160 &  1.48 & ~~0.663 & -0.033 &  58 \\
 92.03 &  H 41 $\alpha$      &  45  &  ~$-$80, 80 &  0.42 & ~~0.670 & -0.036 &  66 \\
 92.49 &  $^{13}$CS          &  44  & $-$100, 140 &  0.70 & ~~0.666 & -0.033 &  53 \\
 93.17 &  N$_{2}$H$^{+}$     &  52  & $-$200, 200 &  2.63 & 359.882 & -0.074 &  14 \\
\hline
\end{tabular}
\end{center}
* = line wing coincides with bad data at sub-band edge.
\end{table*}

\subsection{Peak emission images}
\label{subsec:peak_em}

\begin{figure*}
\includegraphics[angle=-90,width=16.8cm]{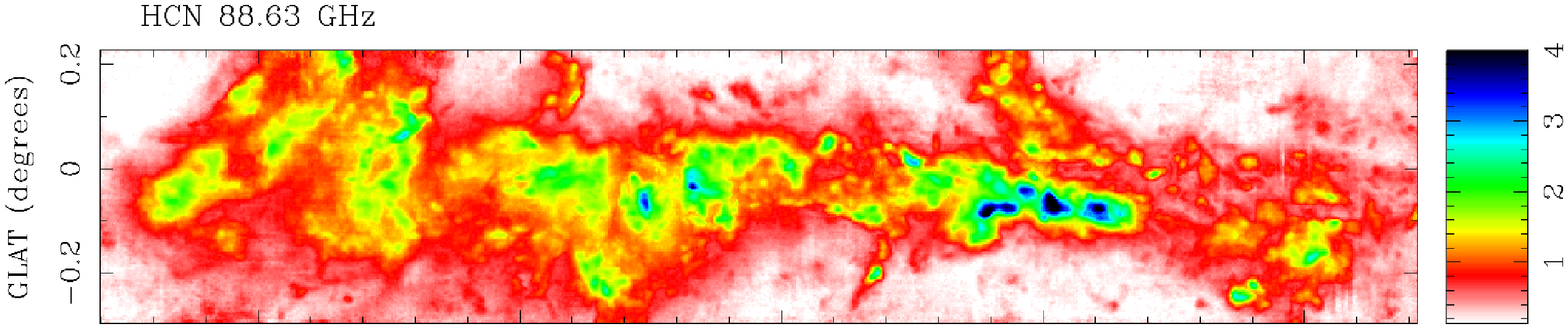}
\includegraphics[angle=-90,width=16.8cm]{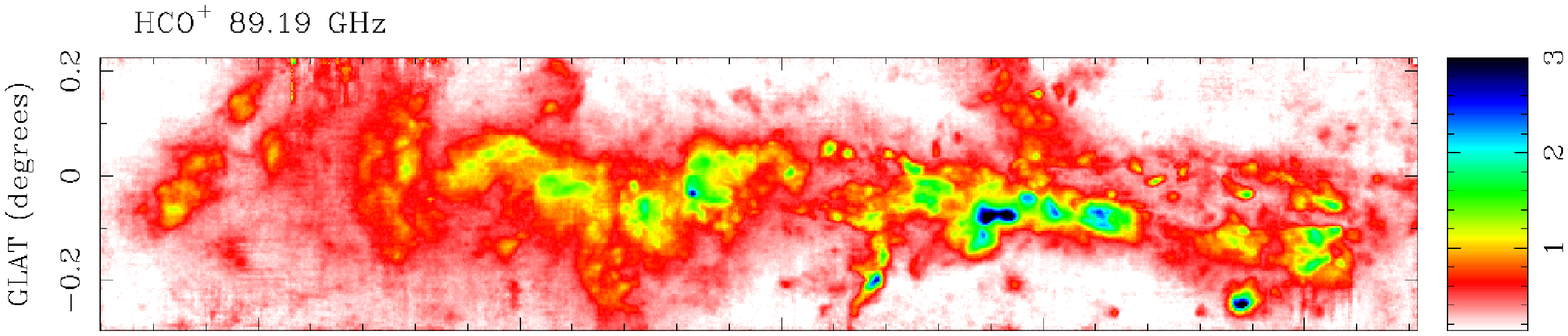}
\includegraphics[angle=-90,width=16.8cm]{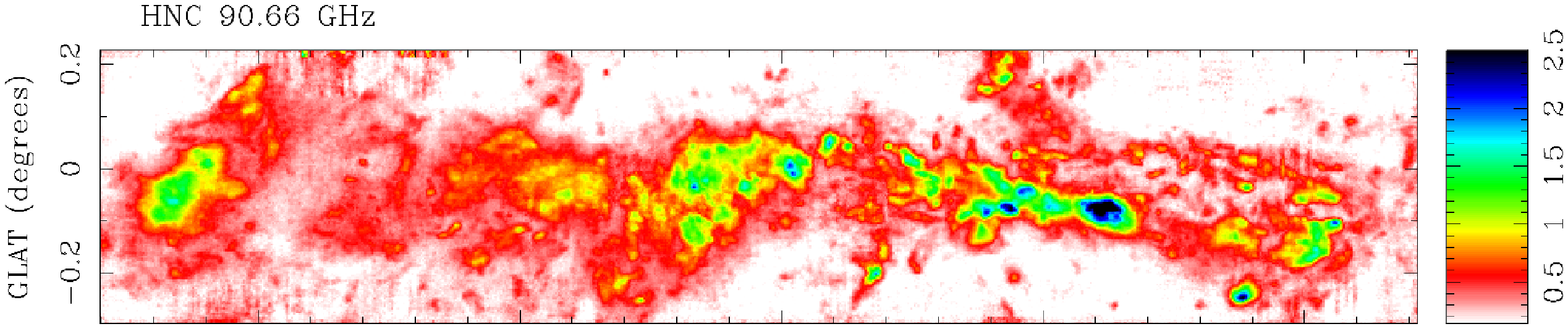}
\includegraphics[angle=-90,width=16.8cm]{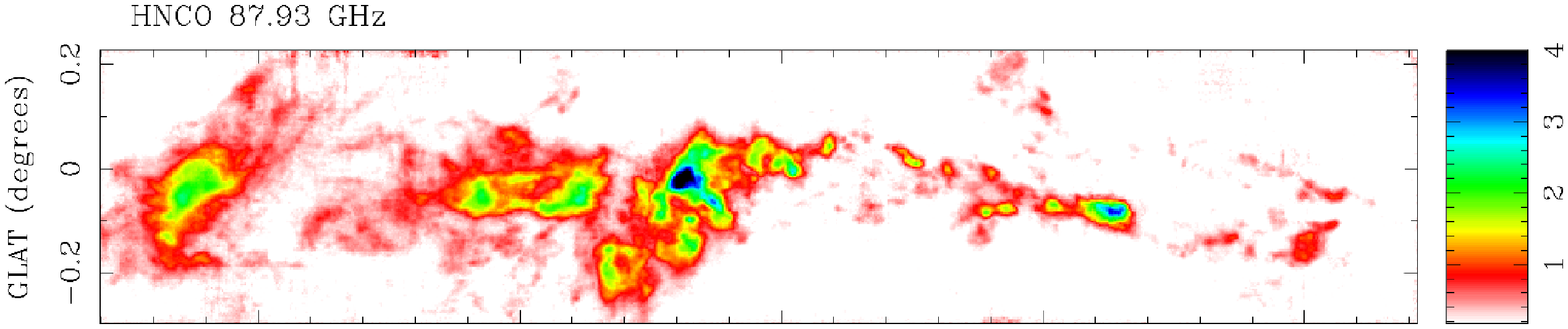}
\includegraphics[angle=-90,width=16.8cm]{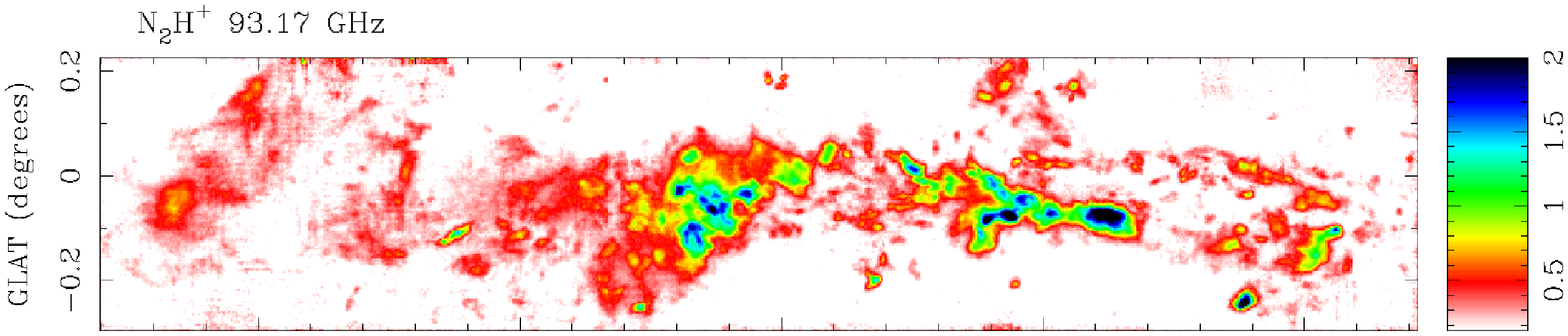}
\includegraphics[angle=-90,width=16.8cm]{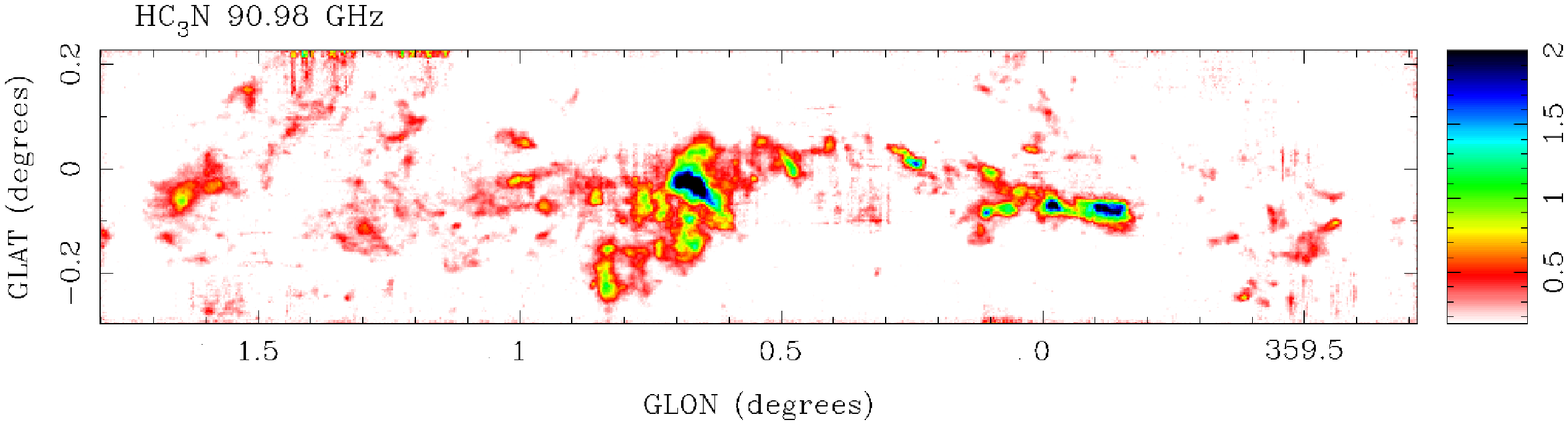}
\caption{Peak brightness images, for the lines of HCN, HCO$^{+}$, HNC,
HNCO, N$_{2}$H$^{+}$ and HC$_{3}$N. These are the six brightest lines we 
observed.
The grey-scale is peak brightness as $T_{A}^{*}$ in K here and in Figs
\ref{fig:images_H13CN_etc} to \ref{fig:images_RRL}.}
\label{fig:images_HCN_etc}
\end{figure*}

\begin{figure*}
\includegraphics[angle=-90,width=16.8cm]{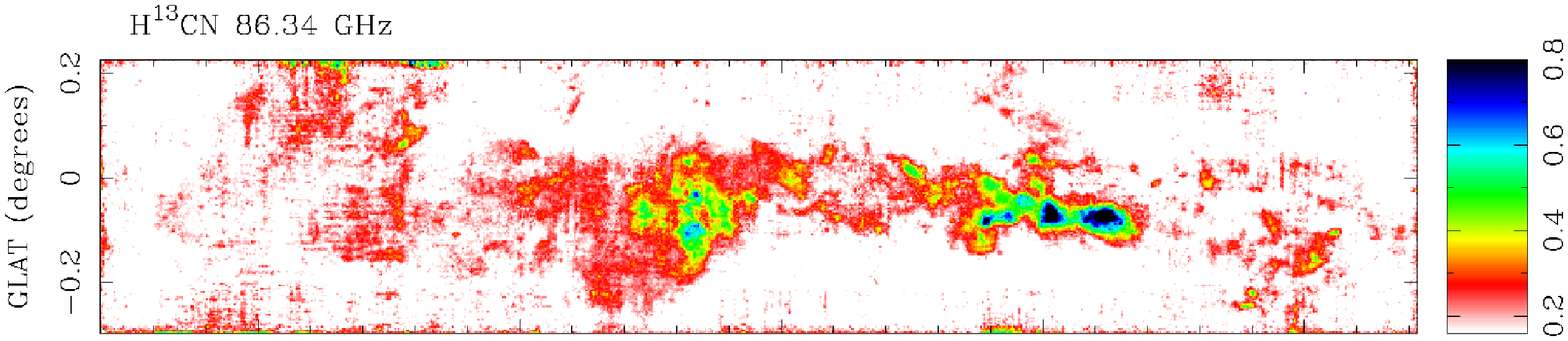}
\includegraphics[angle=-90,width=16.8cm]{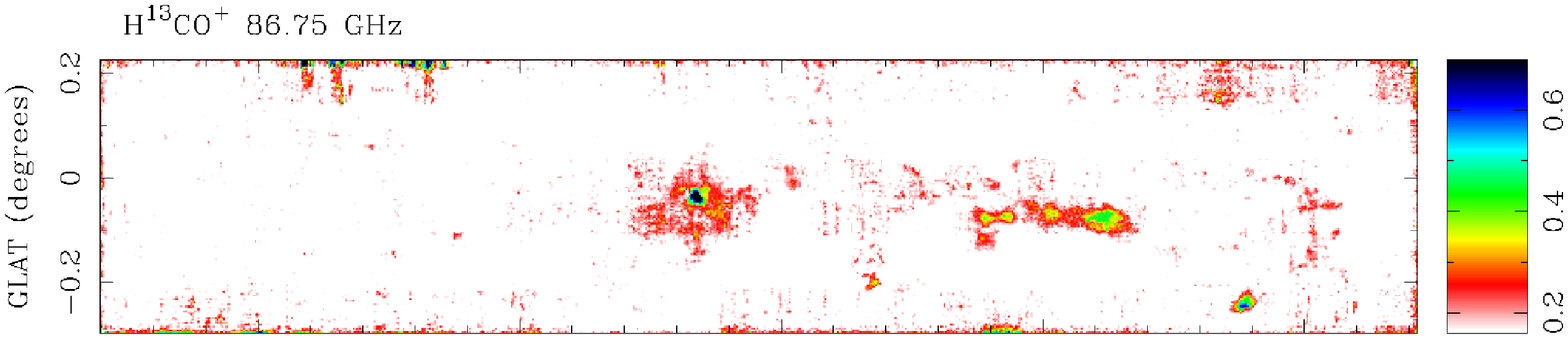}
\includegraphics[angle=-90,width=16.8cm]{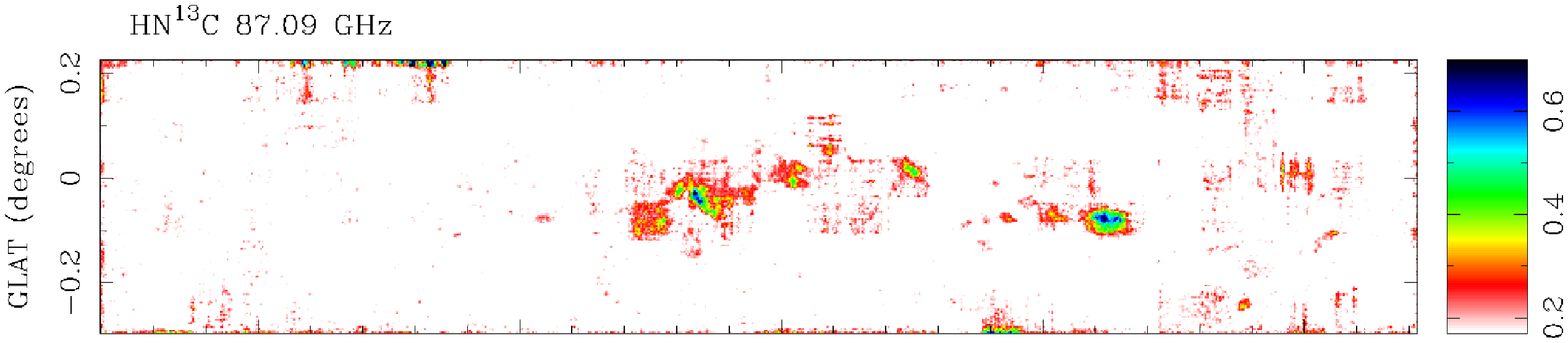}
\includegraphics[angle=-90,width=16.8cm]{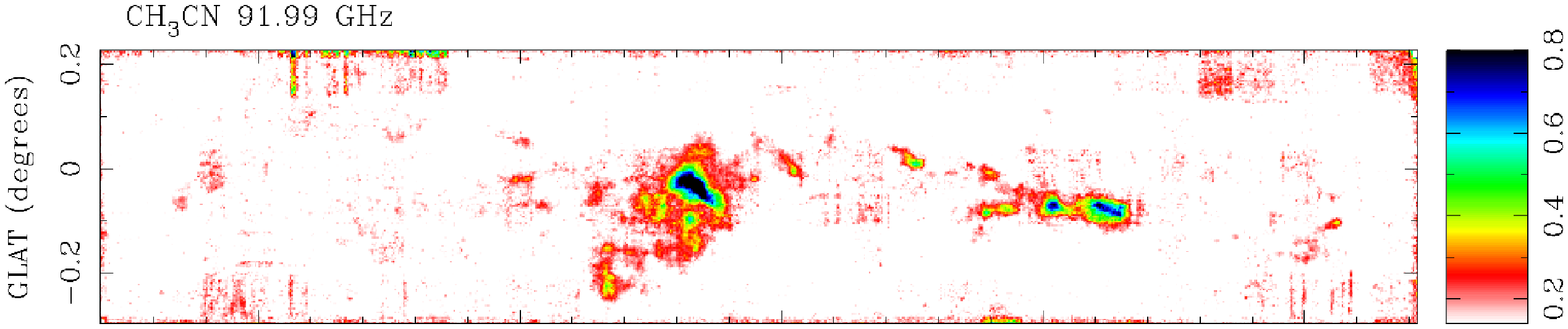}
\includegraphics[angle=-90,width=16.8cm]{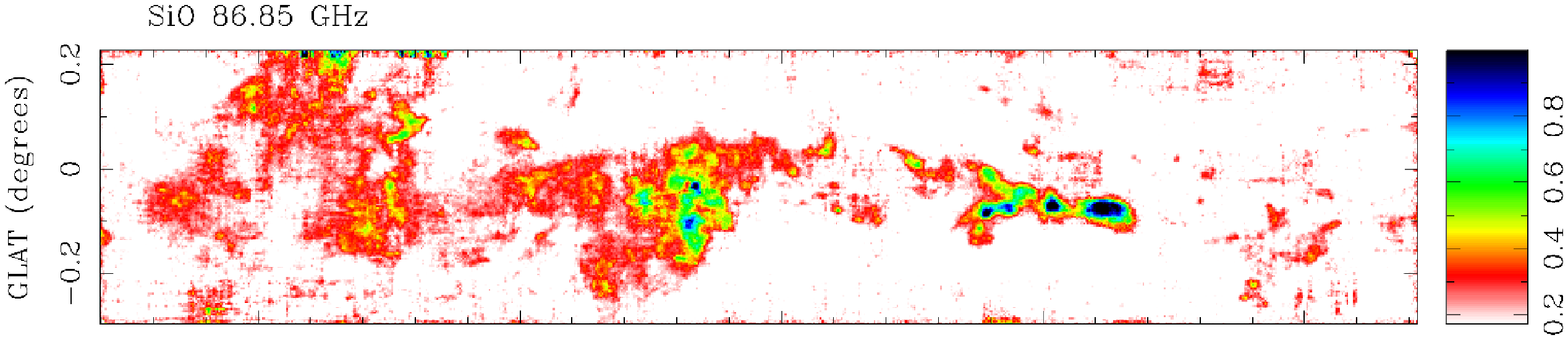}
\includegraphics[angle=-90,width=16.8cm]{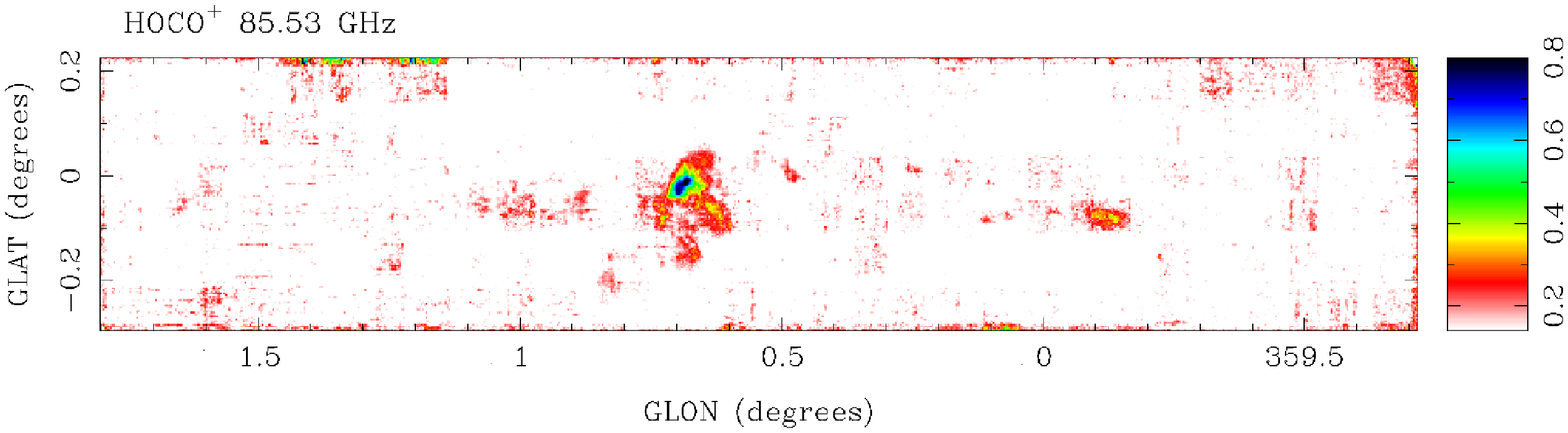}
\caption{Peak brightness images, for the lines of H$^{13}$CN, 
H$^{13}$CO$^{+}$, HN$^{13}$C, CH$_{3}$CN, SiO and HOCO$^{+}$.}
\label{fig:images_H13CN_etc}
\end{figure*}

\begin{figure*}
\includegraphics[angle=-90,width=16.8cm]{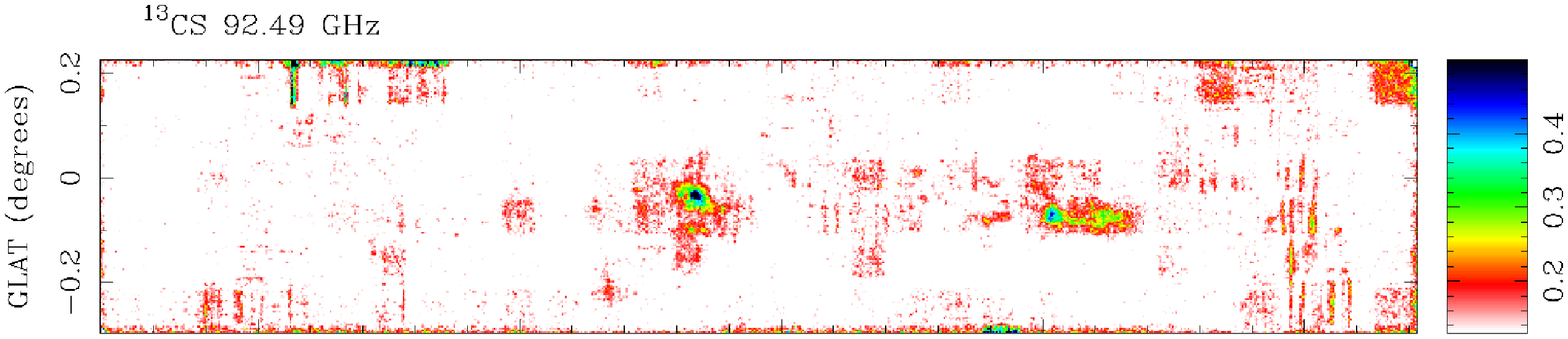}
\includegraphics[angle=-90,width=16.8cm]{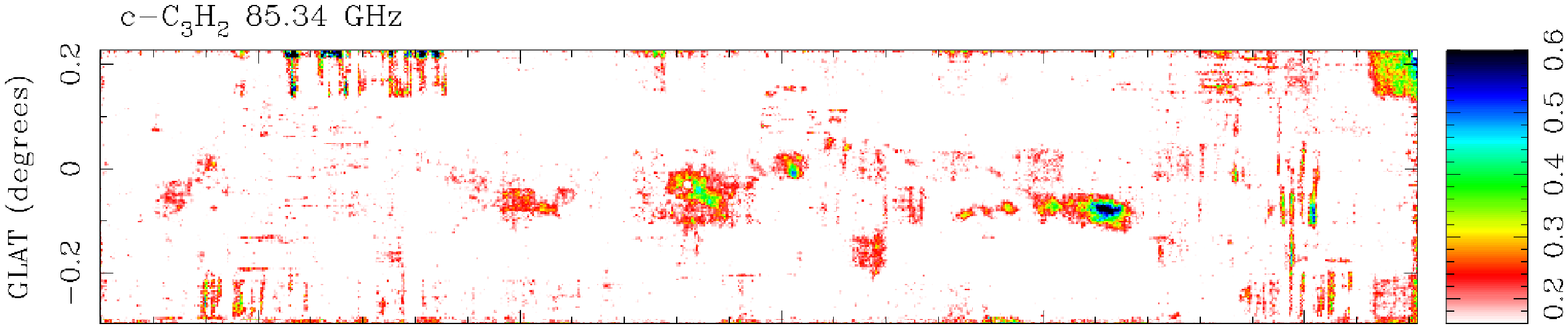}
\includegraphics[angle=-90,width=16.8cm]{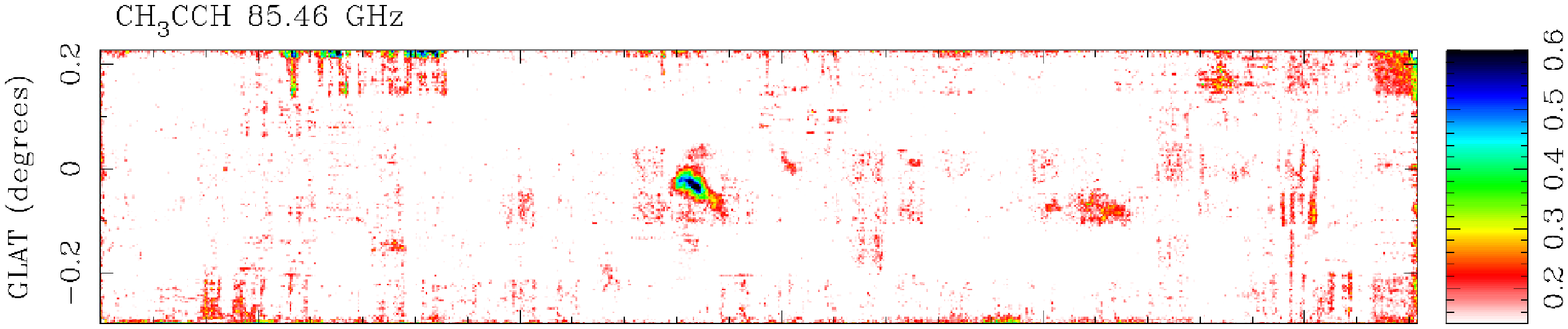}
\includegraphics[angle=-90,width=16.8cm]{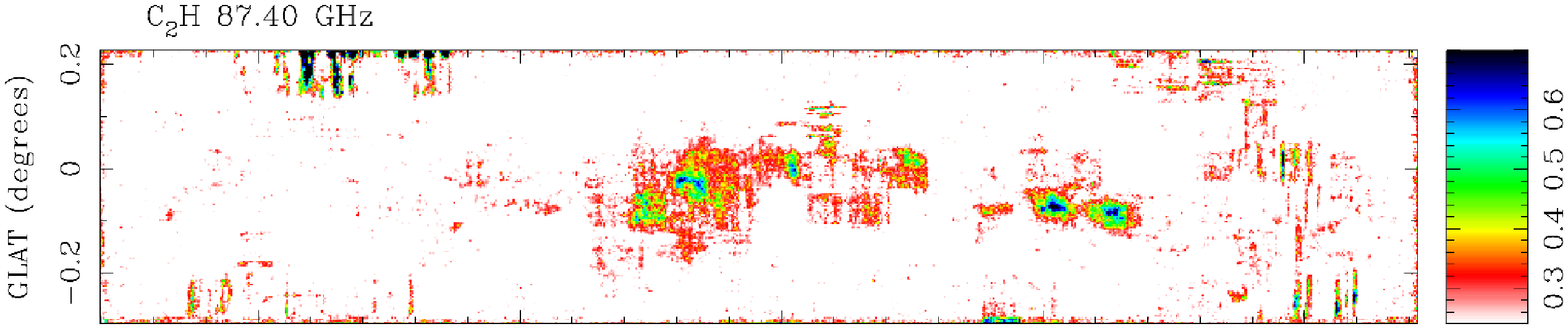}
\includegraphics[angle=-90,width=16.8cm]{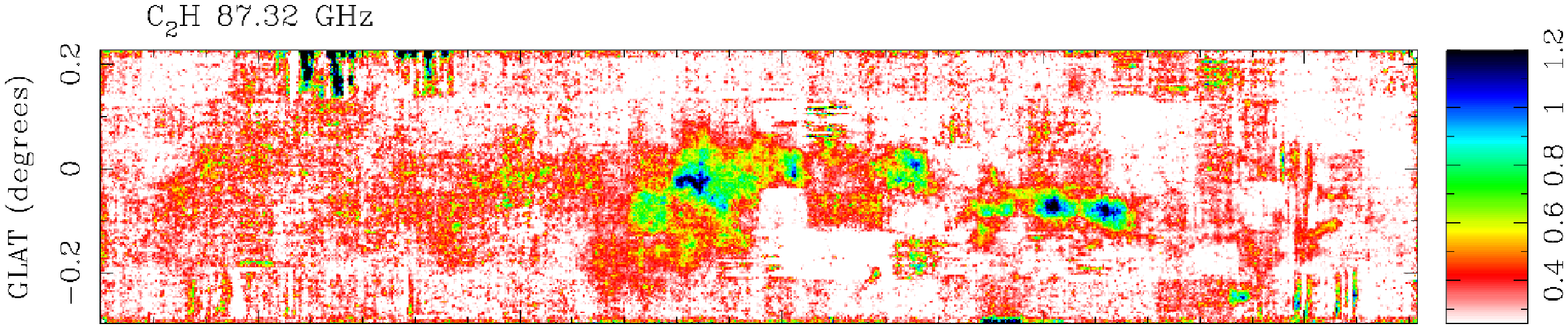}
\includegraphics[angle=-90,width=16.8cm]{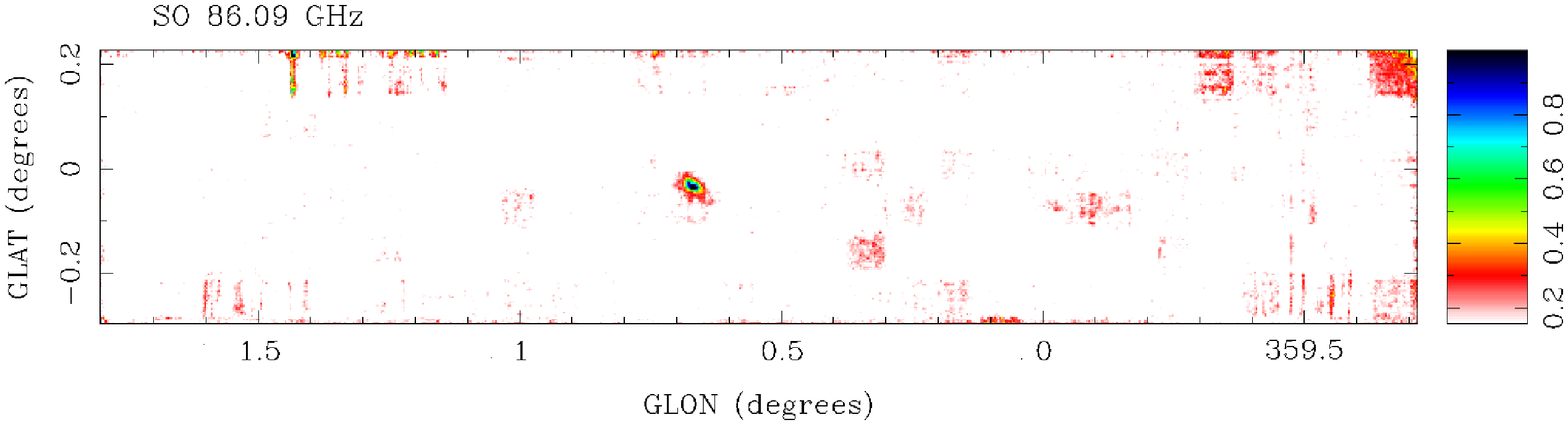}
\caption{Peak brightness images, for the lines of 
$^{13}$CS, c-C$_{3}$H$_{2}$, CH$_{3}$CCH, C$_{2}$H and SO.}
\label{fig:images_13CS_etc}
\end{figure*}

\begin{figure*}
\includegraphics[angle=-90,width=16.8cm]{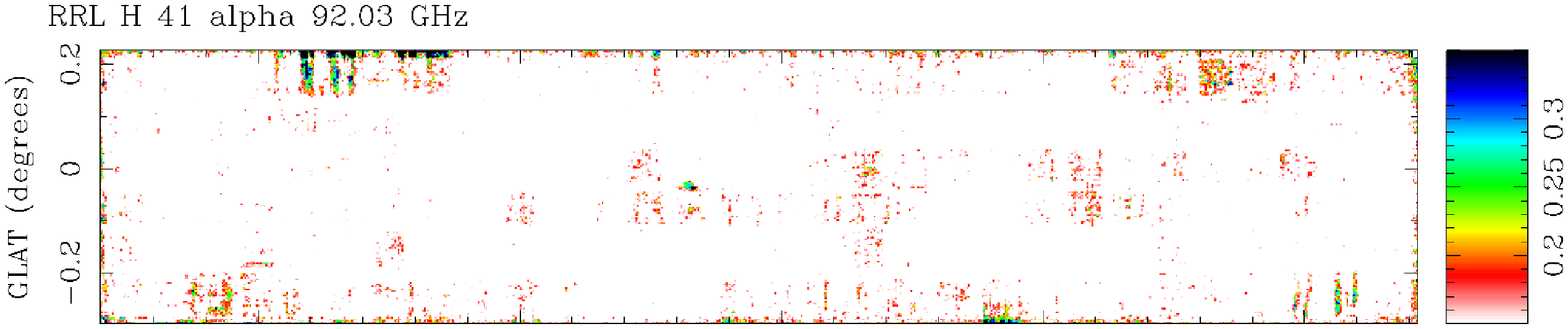}
\includegraphics[angle=-90,width=16.8cm]{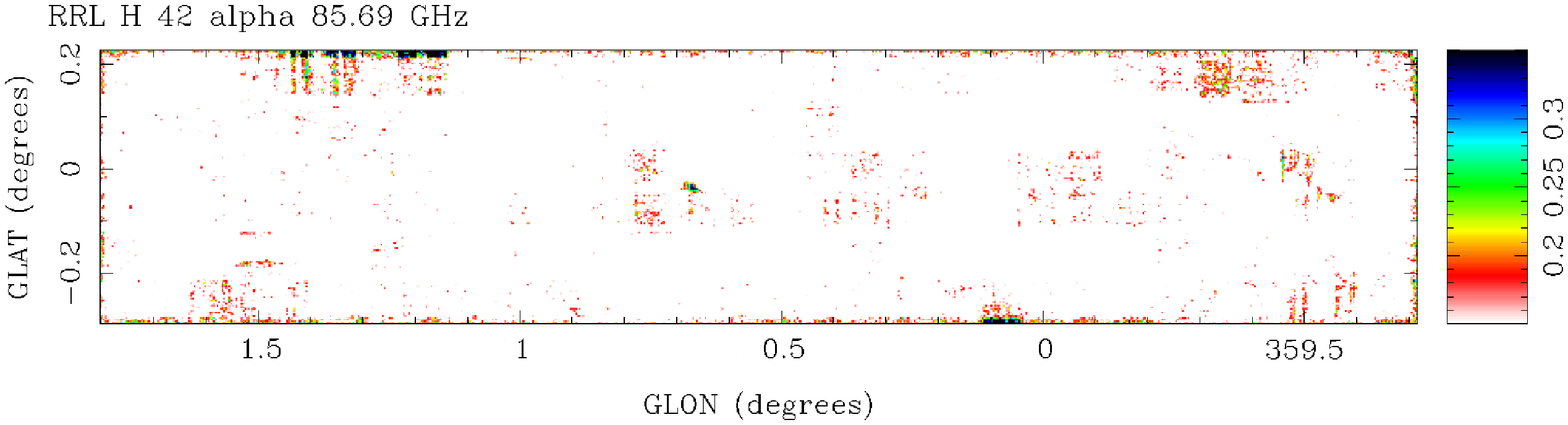}
\caption{Peak brightness images, for the H recombination lines.}
\label{fig:images_RRL}
\end{figure*}

We show the 
2-dimensional distribution of the line emission (as listed in Table
\ref{tab:lines_table}, with quantum numbers) using images of the peak 
emission in Figs \ref{fig:images_HCN_etc} to
\ref{fig:images_RRL}. 
These images are much more robust to the effect of low level baseline problems
than the integrated emission images. 
They also have the advantage of showing fine structure, being less affected
by multiple velocity components.
We use the peak brightness within the
velocity ranges listed in Table \ref{tab:summary_table}.
The grey-scale displays start at the 3 $\sigma$
brightness level, so that the display suppresses areas without significant 
line emission. 
However, it should be noted that there is emission outside the areas 
highlighted in these plots, below the  3 $\sigma$ level for individual pixels,
which is found to be significant 
when the data are integrated over larger areas than the beam. That is, when 
averaging over larger areas, the noise is reduced in terms of brightness,
so that weak emission is then clearer above the reduced noise.

We also note that there are some artifacts in some of these peak brightness 
images, notably stripes in the latitude scanning direction due to data taken
in the poorest weather, around $l = 1.4$~deg., $b = 0.2$~deg. and 
$l = 359.5$~deg., $b = -0.1$~deg.

The peak emission images do not show the line absorption. There is absorption
in some stronger lines (e.g HCN, HCO$^{+}$ and HNC) at the strong continuum of
Sgr B2 \citep{jo+08} and over a wider area, see
subsection \ref{subsec:vel} and Fig. \ref{fig:HCN_pos_vel}. However, 
the peak emission is mostly not sensitive to this absorption, as the velocities
of peak emission are offset from that of the absorbing clouds.

The lines of HCN, HCO$^{+}$ and HNC (Fig. \ref{fig:images_HCN_etc}) are strong
and detected over most of the imaged area. The $^{13}$C isotopologues,
H$^{13}$CN, H$^{13}$CO$^{+}$ and HN$^{13}$C (Fig. \ref{fig:images_H13CN_etc})
are detected in the densest cores, mostly around Sgr~B2 and Sgr~A. The 
ratio between these and the HCN, HCO$^{+}$ and HNC lines shows that the 
latter are optically thick in the densest cores (subsection 
\ref{subsec:line_intensities_ratios}). This optical depth effect explains
why the position of the brightest peak of HCN, and the velocity of the 
brightest peak of H$^{13}$CN are different to that of other lines.

Other strong lines that are detected over a large 
fraction of the CMZ are HNCO, N$_{2}$H$^{+}$,
HC$_{3}$N (Fig. \ref{fig:images_HCN_etc}), CH$_{3}$CN 
and SiO (Fig. \ref{fig:images_H13CN_etc}). The HNCO and HC$_{3}$N peaks are 
particularly strong
around Sgr~B2 (Table \ref{tab:summary_table}).

Weaker lines that are detected 
mostly around Sgr~B2 and Sgr~A are HOCO$^{+}$, 
$^{13}$CS, c-C$_{3}$H$_{2}$,CH$_{3}$CCH  and C$_{2}$H 
(Fig. \ref{fig:images_13CS_etc}). The data for the two lines of the C$_{2}$H 
molecule show noise and greater artifacts due to the line frequencies being 
at the edge of the sub-band, giving poor sensitivity and baseline stability.

The line of SO (Fig. \ref{fig:images_13CS_etc}) is concentrated at 
Sgr~B2, as are the two radio recombination lines, H 41 $\alpha$ and 
H 42 $\alpha$ (Fig. \ref{fig:images_RRL}). These recombination lines
indicate significant free-free emission at 3 mm at Sgr~B2, as the cascade
through Rydberg states of H of recombining electrons occurs under similar
conditions to the thermal bremsstrahlung of the free electrons in the ionised 
gas \citep*{wirohu09}. The free-free emission component of the Sgr~B2 spectrum
at 3 mm is also shown by the overall spectral energy distribution (SED) in
\citet{jo+11}.

The area around Sgr B2 has interesting differences in the line
distributions, which are not well shown at the scale of 
Figs \ref{fig:images_HCN_etc} to \ref{fig:images_RRL}, but which are shown
and discussed in the previous paper \citet{jo+08}.

\subsection{Velocity structure}
\label{subsec:vel}

\begin{figure*}
\includegraphics[width=17.0cm]{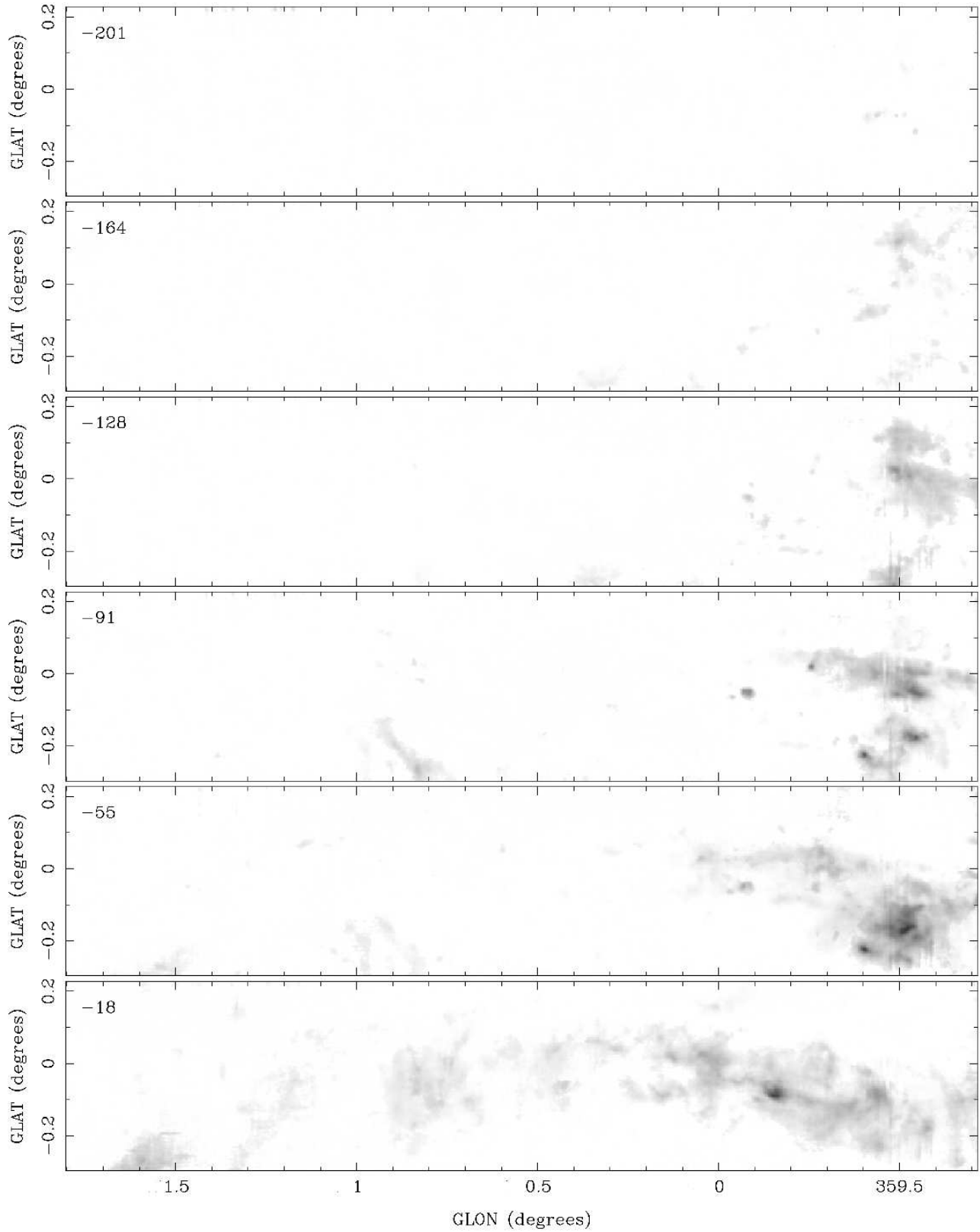}
\caption{The velocity structure in the CMZ as shown by the HCN line channel
images, in 12 panels over 2 pages, averaged over 36~km~s$^{-1}$ ranges, with
the central velocity noted in the top left of each panel. The greyscale range is 
0 to 2~K in T$_{A}^*$.}
\label{fig:HCN_chans}
\end{figure*}

The line emission in the CMZ spans a large range of velocities (Fig. 
\ref{fig:HCN_spec}), which, as we note in subsection \ref{subsec:cubes},
is due to both the large gradient in velocity across the CMZ in the deep 
potential well of the Galactic Centre and large line widths in
the CMZ. We show in Fig. \ref{fig:HCN_chans} images of the HCN line 
emission as a function of velocity, integrated over 36~km~s$^{-1}$ ranges.
This demonstrates the large velocity gradient across the Galactic Centre,
with negative velocities at more negative longitudes, and positive velocities
at more positive longitudes. 

The molecular line emission in the CMZ is not
symmetrically distributed around the Galactic Centre, so the strongest emission
appears at positive longitudes (dominated by Sgr~B2 at around longitude $0.7$
degrees) giving an asymmetric distribution in velocity 
(about $V_{LSR} = 0$~km~s$^{-1}$) in
Fig. \ref{fig:HCN_chans} and the profile in 
Fig. \ref{fig:HCN_spec}. The area we have 
observed to cover the strongest emission is longitude $-0.7$ to $1.8$ deg.,
so there is also some observational selection in this bias to positive 
longitudes and velocities.

Another way to show the velocity structure is with a position-velocity
diagram, as in Fig. \ref{fig:HCN_pos_vel} for the peak brightness of HCN
as a function of velocity and longitude, within three latitude ranges 10~arcmin 
thick
covering -0.29 to 0.21~deg. This shows both the velocity gradient 
across the region, as well as large velocity widths at some locations.

Also shown in Fig. \ref{fig:HCN_pos_vel} are absorption features at velocities
around $-52$, $-28$ and $-3$~km~s$^{-1}$, which are also seen in
the integrated emission spectrum (Fig.\ref{fig:HCN_spec}). These absorption 
features are clear in the channel images of the HCN data cube, but are not 
obvious in Fig. \ref{fig:HCN_chans} because of the large velocity range
(36~km~s$^{-1}$) summed in this plot.
Considering the overall Galactic rotation with roughly circular motions around
the Galactic Centre, gas along the line of sight to the Galactic Centre
will be moving transverse to the line of sight, with zero radial velocity. 
However, the motions within the Galaxy are not exactly circular, and the 
absorption features at different velocities are associated with intervening 
Galactic features \citep{grwi94, wi+10}. The lines are in absorption, as this 
intervening low density gas has lower excitation temperature than
the background brightness temperature, with excitation temperature $T_{ex}$
of a few K found for CS in sightlines to Sgr~B2 by \citet{grwi94}.

We show in Fig. \ref{fig:peak_vel} the image of the peak velocity from 
N$_{2}$H$^{+}$ emission data cube.  
This line generally shows a well defined peak (see also 
Fig. \ref{fig:misc_profiles}) and so provides a representation of the principal 
kinematic velocity at each position in the CMZ.  The bulk of the emission peaks 
at $V_{LSR} \sim +50$~km~s$^{-1}$, however there is a clear gradient in 
Galactic longitude of $\sim 200$~km~s$^{-1}$ evident, extending from 
$\sim +100$~km~s$^{-1}$ at $l=+1.5^{\circ}$ to $\sim -100$~km~s$^{-1}$ at 
$l=-0.5^{\circ}$.  More complex kinematics is also evident.  In particular, 
the twisted cold dust ring noted by \citet{mo+11}  is apparent in the 
N$_2$H$^+$ peak velocity image.   This elliptical dust ring, 
$\sim 100 \times 60$\,pc across, has been hypothesised to trace the location 
of the system of stable $x_2$ orbits in the barred Galactic potential.  
It is apparent in a dust column density image \citep[see fig. 4][]{mo+11}, and 
shows a coherent velocity structure as traced in the emission range of the 
CS 1-0 line \citep{tshauk99}.
The kinematics of this ring are more clearly shown in the N$_2$H$^+$ peak 
velocity image shown in Fig. \ref{fig:peak_vel}, however.  
At $l=-0.5^{\circ}$ the western end of the 
ring is most clearly seen, with velocities smoothly changing from $-20$ to 
$-100$~km~s$^{-1}$ around the ring's extremity. The twist in the ring occurs 
around $l \sim +0.1^{\circ}$ as it crosses over itself.  The eastern side, 
around $l \sim 0.5 - 0.8^{\circ}$, is less readily identified in the velocity 
images, but the emission velocity nevertheless varies smoothly across it, 
from $20 - 50$ km~s$^{-1}$ across the centre of the ring, to 
$\sim +100$~km~s$^{-1}$ at its eastern end.  The ring twists 
about the Galactic plane, rising above and below it twice, for one complete 
orbit. \citet{mo+11} propose that this ring is rotating about the centre 
of the Galaxy with an orbital velocity of 80~km~s$^{-1}$, tracing the stable, 
non-intersecting $x_2$ orbits whose major axis is perpendicular to that of the 
Galactic bar, and precesses as the the bar does.

\begin{figure*}
\includegraphics[width=17.0cm]{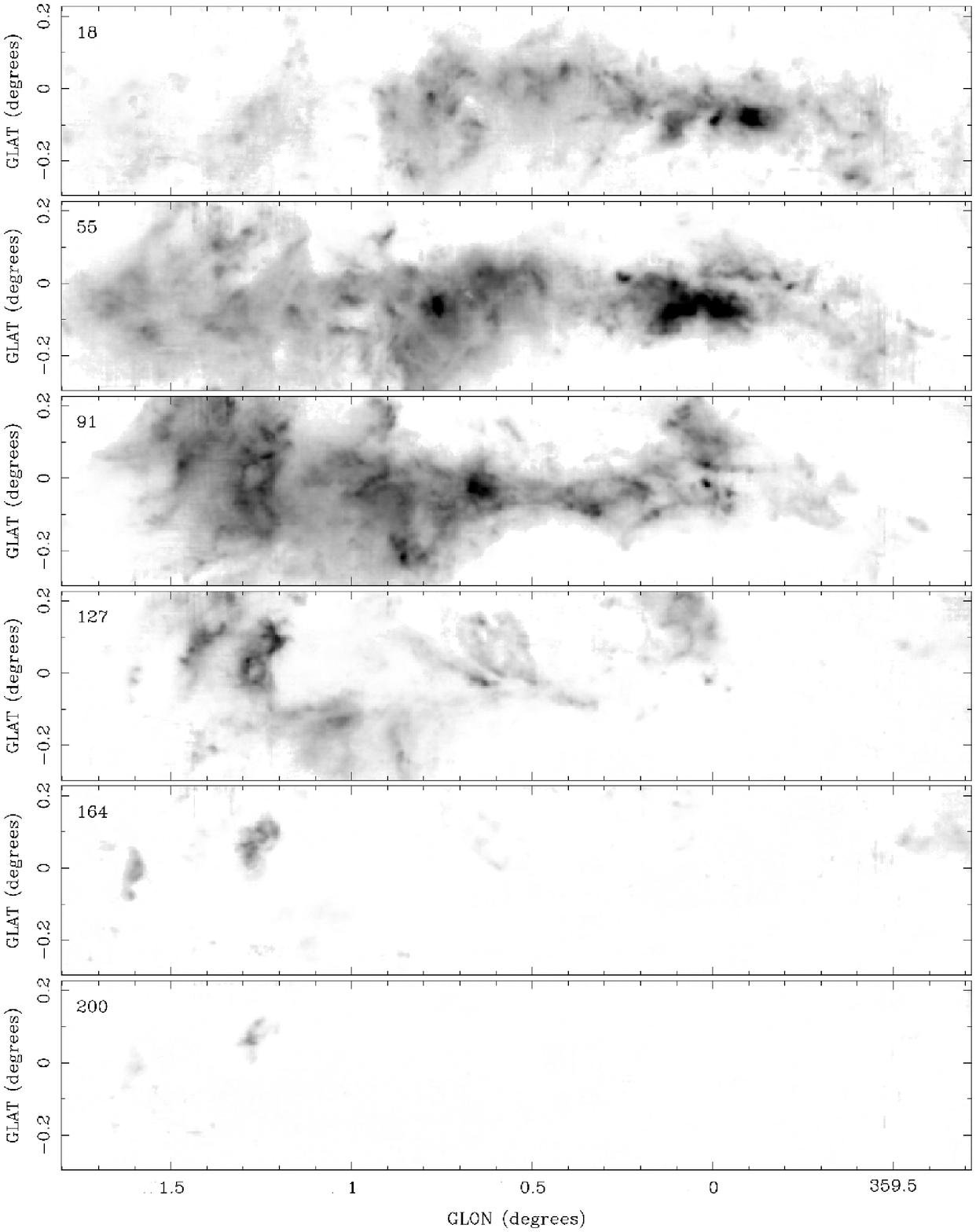}
{\bf Figure \ref{fig:HCN_chans}} (continued)
\end{figure*}

\begin{figure*}
\includegraphics[angle=-90,width=17.0cm]{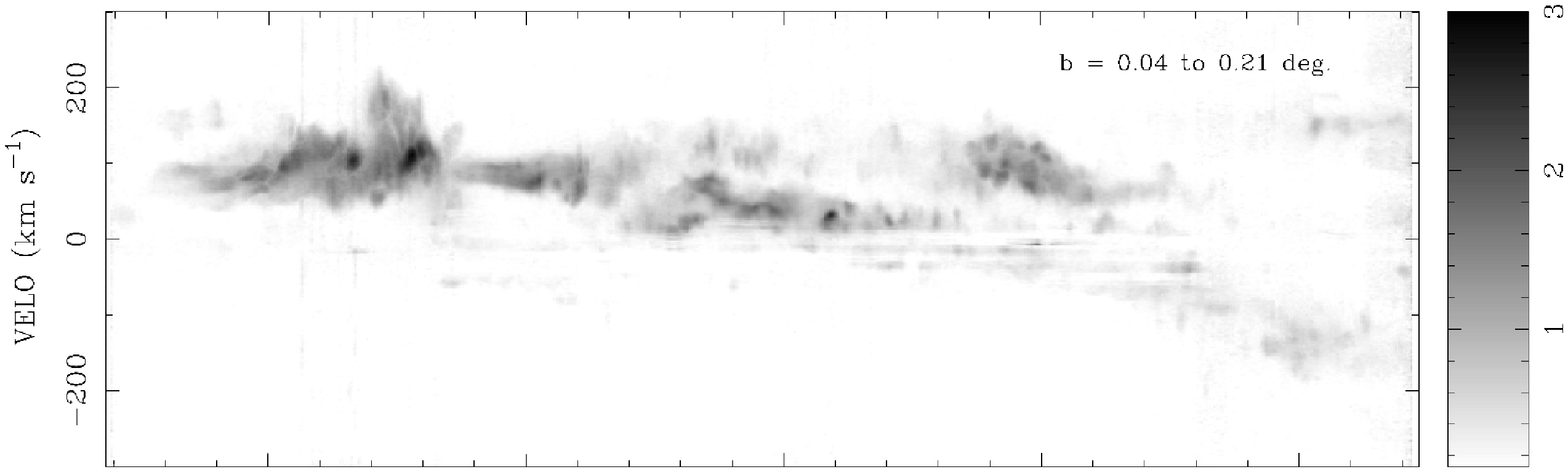}
\includegraphics[angle=-90,width=17.0cm]{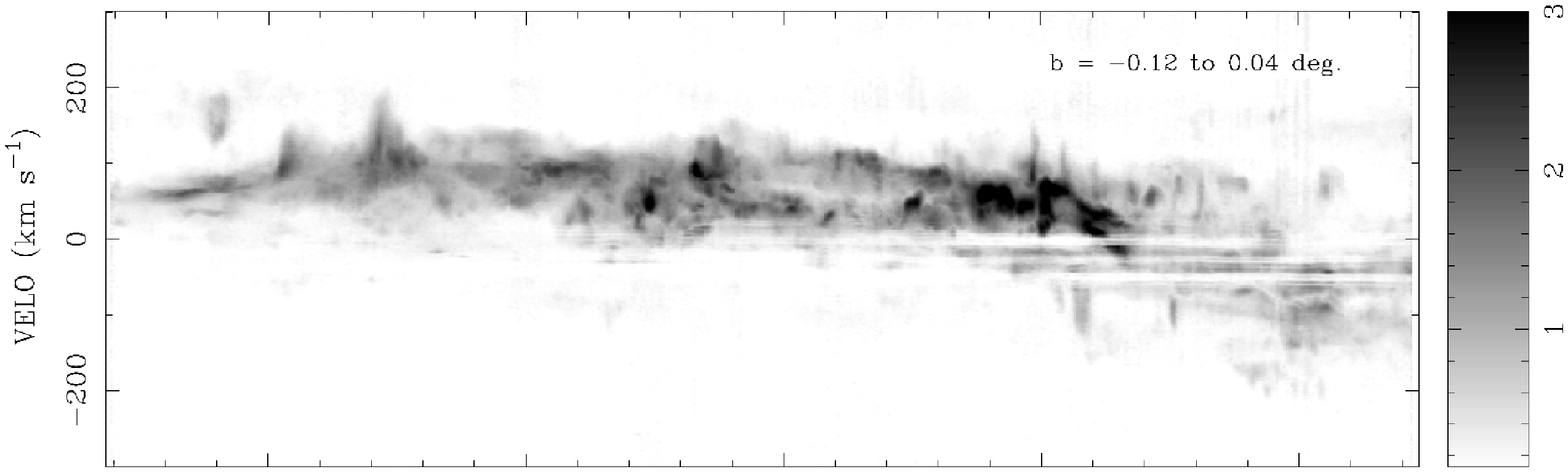}
\includegraphics[angle=-90,width=17.0cm]{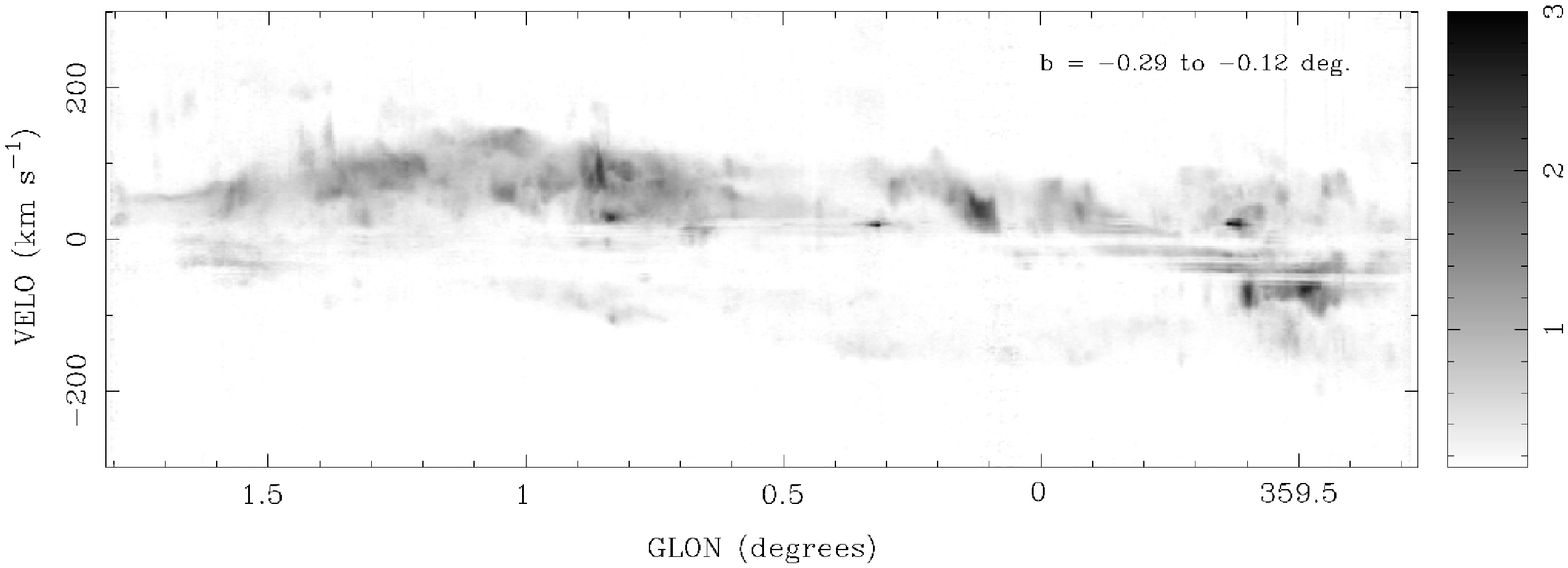}
\caption{Position-velocity diagram of the CMZ, from HCN data, shown as peak
brightness temperature, within the three latitude ranges -0.29 to -0.12~deg.
(bottom), -0.12 to 0.04~deg. (middle) and 0.04 to 0.21~deg. (top).
Note 
the velocity gradient, and the absorption features at around $-52$, $-28$
and $-3$~km~s$^{-1}$.}
\label{fig:HCN_pos_vel}
\end{figure*}

\begin{figure*}
\includegraphics[angle=-90,width=17.0cm]{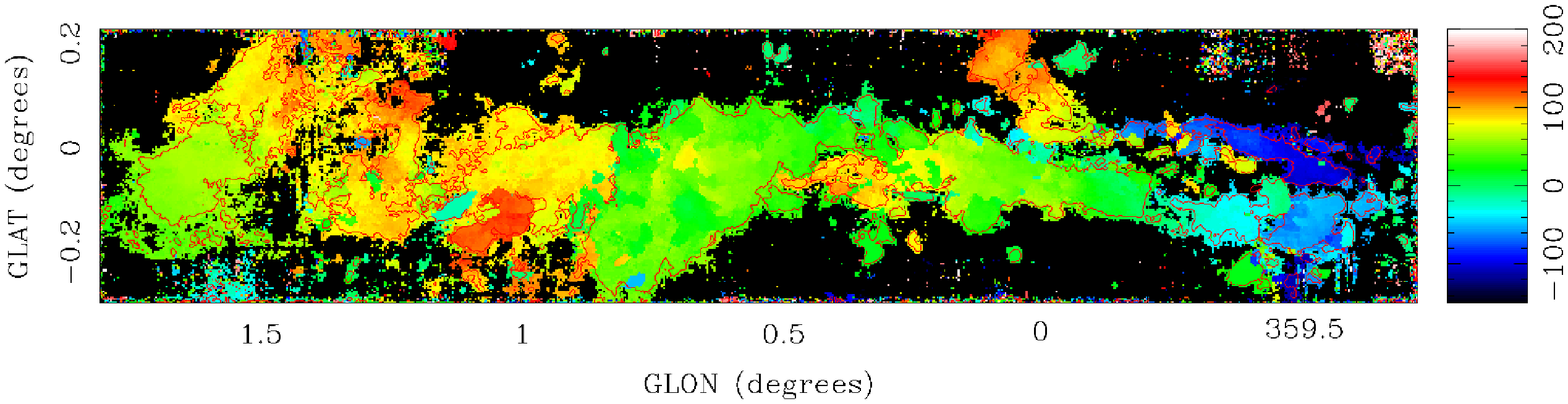}
\caption{The peak velocity image, from the N$_{2}$H$^{+}$ line, with the scale in 
km~s$^{-1}$, showing the overall velocity gradient in longitude, but also more
complex kinematics. The 0.25 K contour from N$_{2}$H$^{+}$ peak brightness is 
plotted as a guide to the features.}
\label{fig:peak_vel}
\end{figure*}

\subsection{Integrated line emission}
\label{subsec:integ_em}

For quantitative analysis of the line data cubes, it is useful to integrate
the emission over the velocity range. As the velocity range is large (Table 
\ref{tab:summary_table} and subsection \ref{subsec:vel} above), the integrated
emission images are sensitive to noise and problems with the spectral baselines,
as noted
in subsection \ref{subsec:cubes}. This is a well known problem in the analysis
of spectral-line data, as discussed for example by \cite{vaek89} and 
\citet{ru99}.

The major problem of simply integrating the emission is spurious stripes, in 
the longitude and latitude scanning directions,
caused by spectral baseline ripples, dominated by the small fraction of 
data taken during the worst weather. A further problem is that extra noise is 
added from including parts of the spectra where there is no signal.  

A simple method that we adopt here is to integrate the emission above a fixed 
cutoff level at which the emision is significant. We use a 3-$\sigma$ cutoff,
the same used in the peak emission plots (Figs. \ref{fig:images_HCN_etc} to 
\ref{fig:images_RRL}) to define significant emission. This is quite 
effective in removing the spurious features in the integrated emission, 
but not totally effective as the non-gaussian statistics of the baseline errors
means there is still a small tail to the distribution of spurious
features above the 3 $\sigma$ level. Some of these artifacts can be seen in 
Figs. \ref{fig:images_HCN_etc} to \ref{fig:images_RRL}.

However, a major problem with this simple method of integration above a
cutoff level is a bias in the integrated emission \citep{vaek89,ru99},
with real emission below the cutoff being missed. Because of this, we
have investigated two other methods of processing to obtain integrated emission.

An alternative to the fixed cutoff level, is to integrate the emission within
a mask defined by the data smoothed spatially and/or over velocity
\citep{vaek89,ru99}. We have used as the mask a wavelet reconstruction of the
cube, implemented as part of the 
Duchamp\footnote{http://www.atnf.csiro.au/people/Matthew.Whiting/Duchamp/}
source finding package \citep{wh08}. This gives a model spectrum at each pixel
which contains a range of velocity scales, but not the smallest scale, so
has reduced noise, and so can be extended to lower brightness levels than the
original data. This gave quite similar results in integrated emission images
to the 3-$\sigma$ cutoff version, with slighly more integrated emission, but
also slightly more artifacts. It was decided that any improvement was not worth
the extra complexity of the processing.

We have also tested different methods of processing the data cubes, with a spectral 
baseline correction, to give a filtered version, to help the problematic low
level ripples. The baseline was determined by an iterative process
that clips the data above the 3 $\sigma$ level, replaces these clipped pixels 
with zero, and then smooths the data with a boxcar
filter of width 19 pixels (around 34~km~s$^{-1}$). The baselines in parts of the
data cube with significant emission (above the 3 $\sigma$ level) are unaffected.
For parts of the data cube without significant emission, the filtering removes
low level offsets to make the mean zero. This is similar to a high order
polynomial fit to the spectral baselines (after the exclusion of significant
line emission). The baseline corrections are a few times 10 mK,
or around half the noise level in the original cubes (Table 
\ref{tab:summary_table}). In practice, the integrated emission determined
from the filtered cubes (after a 3-$\sigma$ cutoff, with the reduced level
of $\sigma$) was less than that from the unfiltered cubes with the
original 3-$\sigma$ cutoff. That is, the effect of integrating to a lower level
is overwhelmed by the effect of the filtering biasing the spectra down.

In practice, then, we use the simple integration of emission above a fixed  
3-$\sigma$ cutoff level, with the caveat that this will miss some flux from 
low-level emission. This is a tradeoff to remove the obvious artifiacts.
However, since the baseline stability in the data is a limitation,
such very low-level broad emission is not measured accurately
anyway, so this a limit to the original data, rather than just a problem
introduced by the processing. 

The integrated emission images show similar features to
the the peak emission images Figs.
\ref{fig:images_HCN_etc} 
to \ref{fig:images_RRL}, and look to the eye much like the peak emission 
images, but with less contrast due
to overlapping structures, so are not plotted here.

\subsection{Principal Component Analysis}
\label{subsec:pca}

The different spectral lines in the CMZ show similarities as well as
differences (Figs. \ref{fig:images_HCN_etc} to \ref{fig:images_RRL}).
It would be useful to identify and quantify these features of the data, in a 
simple objective manner. One useful technique to do so
is Principal Component Analysis (PCA), see e.g \citet{hesc97}. 

This describes
the multi-dimensional data set by linear combinations of the data that describe
the largest variance (the most significant common feature) and successively
smaller variances (the next most significant features). 

In the context here of integrated emission images,
we can use PCA to describe the images, with a smaller set of
images which contain the most significant features. This has been used by 
\citet{un+97} for the OMC-1 ridge, and more recently by us for the G333 
molecular cloud \citep{lo+09} and the Sgr~B2 area with 3-mm and 7-mm molecular 
lines \citep{jobulo08,jo+11}.

We have implemented the PCA processing in a {\sc python} script, with the 
PCA module\footnote{http://folk.uio.no/henninri/pca\_module/}, and 
pyFITS\footnote{http://www.stsci.edu/resources/software\_hardware/pyfits/} 
to read and write the FITS images.  

Since the PCA finds features in normalised versions
of the different input data sets, it does not work well with low signal to 
noise data, as the noisy data
is scaled up, and the PCA will be dominated by spurious features. So, we 
restrict our PCA analysis to the strongest eight lines (HCN, HCO$^{+}$, HNC, 
HNCO, N$_{2}$H$^{+}$, SiO, CH$_{3}$CN and HC$_{3}$N) out of the total twenty
lines, and an area around Sgr~B2 and Sgr~A with the strongest emission. 
This area was defined by a mask where the  
integrated emission of the N$_{2}$H$^{+}$ line was stronger than 10 
K~km~s$^{-1}$. The normalisation 
of the integrated images makes
versions that are mean zero and variance unity. 

\begin{figure*}
\includegraphics[angle=-90,width=17.0cm]{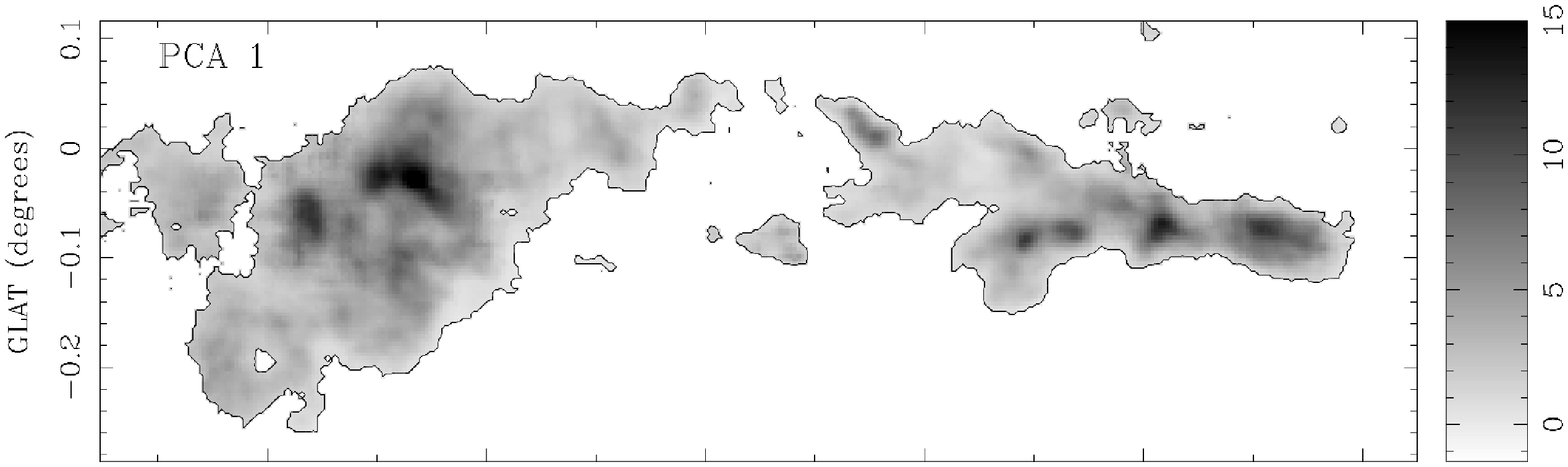}
\includegraphics[angle=-90,width=17.0cm]{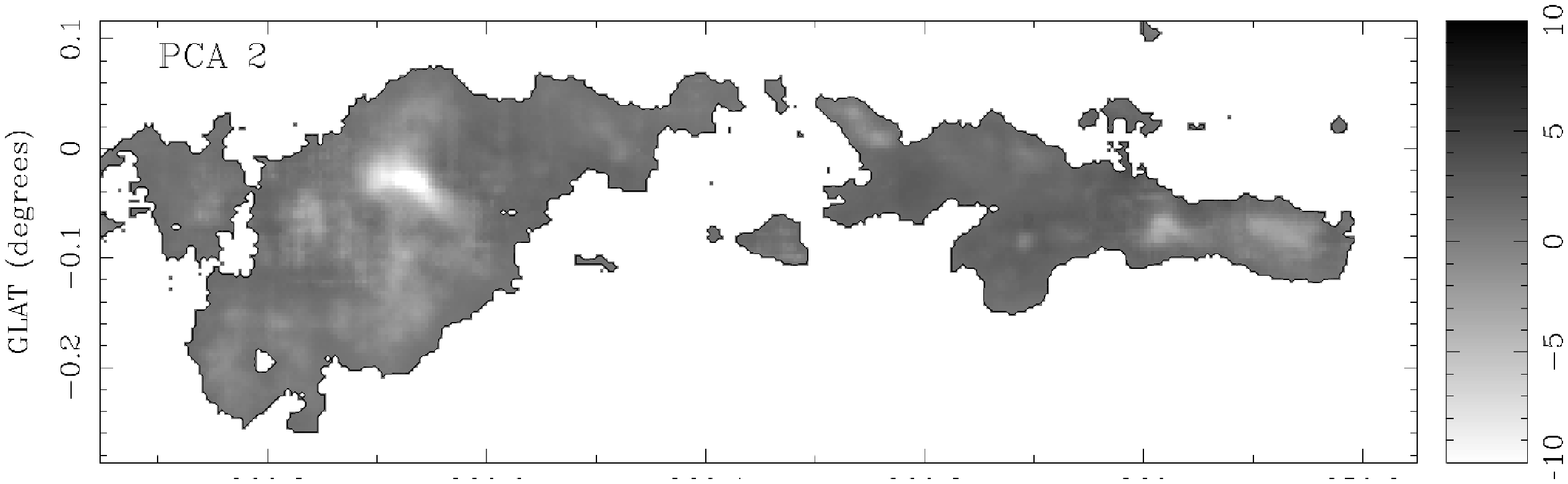}
\includegraphics[angle=-90,width=17.0cm]{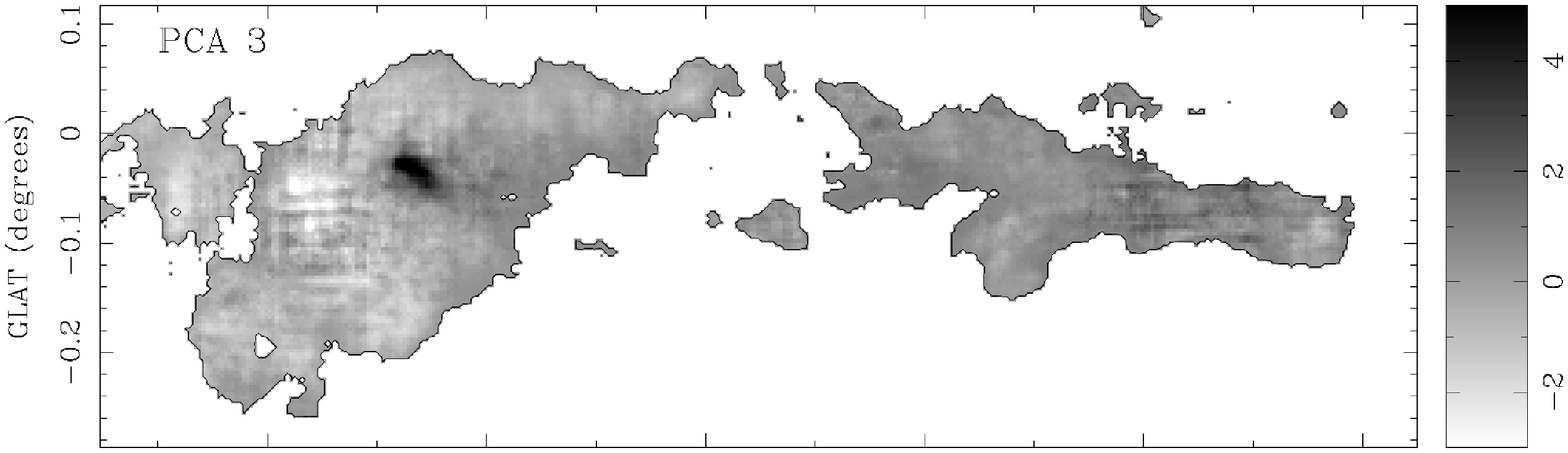}
\includegraphics[angle=-90,width=17.0cm]{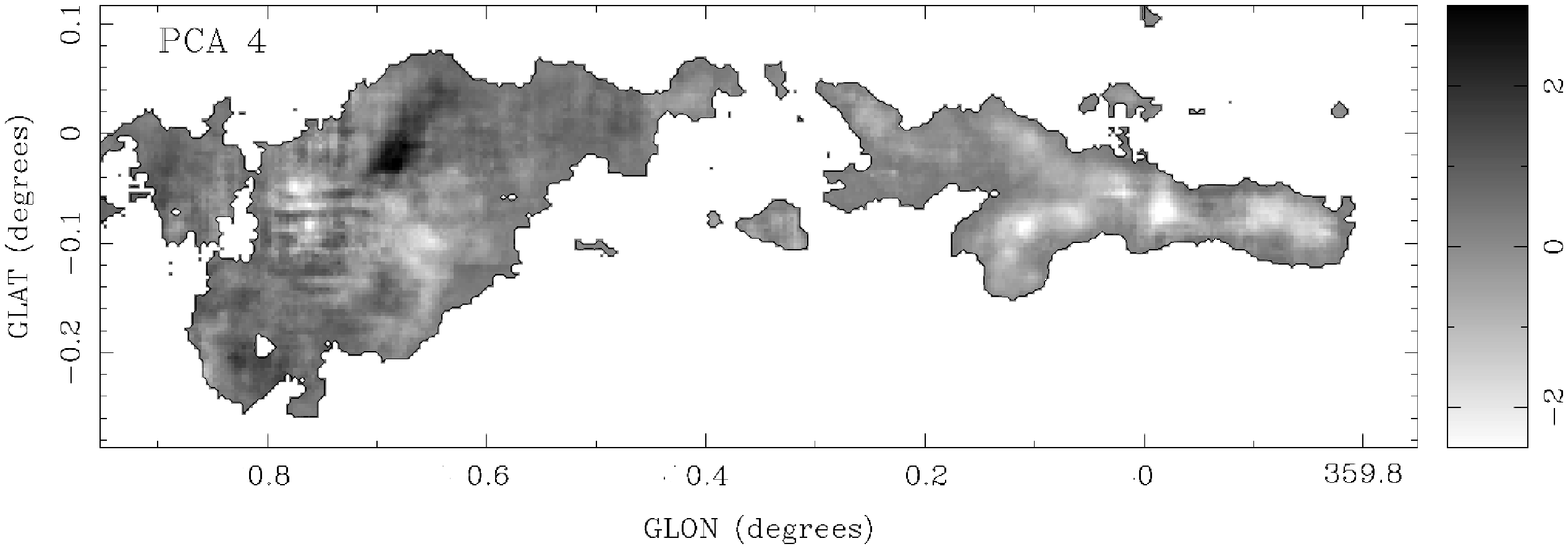}
\caption{The first four principal component images derived from the eight
strongest lines (HCN, HCO$^{+}$, HNC, HNCO, N$_{2}$H$^{+}$, SiO, CH$_{3}$CN
and HC$_{3}$N), which describe 82.6, 10.6, 2.8 and 1.7 percent of
the variance respectively.  The first
principal component describes the common features of the eight lines,
and the second principal component describes the most significant differences
between the lines. The further principal components describe successively
smaller differences. As the PCA requires good signal-to-noise data, we
have restricted the lines used to the brightest and the area to that defined
by the N$_{2}$H$^{+}$ emission (see text), as shown by the contour.}
\label{fig:PCA}
\end{figure*}

\begin{figure*}
\includegraphics[width=5.7cm]{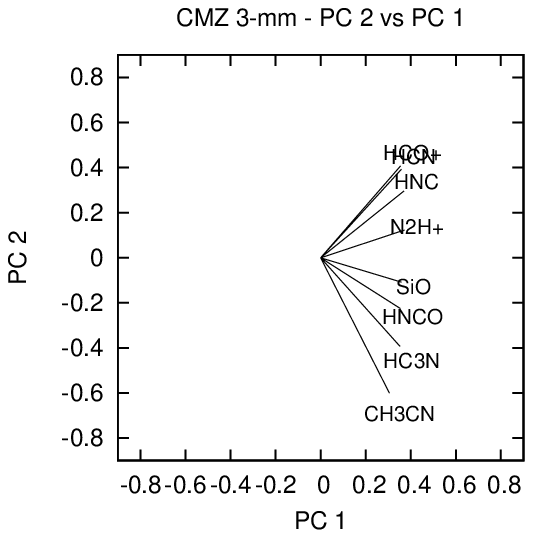}
\includegraphics[width=5.7cm]{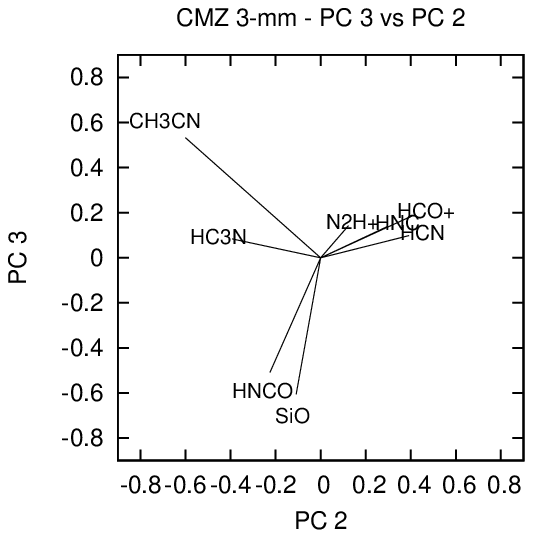}
\includegraphics[width=5.7cm]{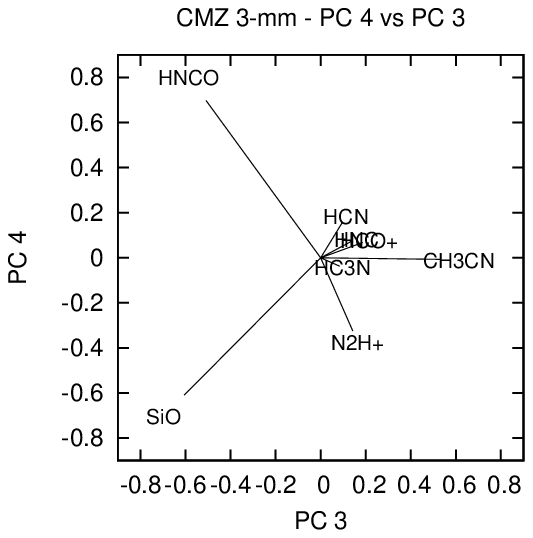}
\caption{The component vectors of the integrated line images in the 
decomposition of the data into the PCA images, shown with successive pairs of 
PCA images. HCN, HCO$^{+}$ and HNC have similar vectors so the labels overlap.
These vectors describe how the normalised integrated images of different lines
are composed of the sum of different scalar amounts of the principal component
images (Fig. \ref{fig:PCA}). For example, all eight lines have distributions
which are made up of positively scaled amounts (PC 1, $\sim$ 0.3) of 
the first principal component,
but are further made up of positive or negative amounts (PC 2) of the
second principal component etc.}
\label{fig:PCA_vect}
\end{figure*}

The images of the four most significant components of the integrated 
emission are shown in Fig. \ref{fig:PCA}. These four components 
describe respectively 82.6, 10.6, 2.8 and 1.7  percent
of the variance in the data.
These components are statistical descriptions of the integrated line images,
not necessarily physical components of the CMZ. However, they do highlight
the physical features in a useful way. 
The normalised projection of the molecules onto the principal
components are shown in Fig. \ref{fig:PCA_vect} with successive pairs of
components.

The first principal component (Fig. \ref{fig:PCA}) shows emission common to
all eight strong lines, or a weighted average integrated emission. It describes
a large fraction (82.6  percent) of the variance, which means that all eight lines
are quite similar in overall morphology. 
The major features are the ridge of emission in Sgr~B2
\citep{jo+08,jobulo08} and the dense cores in Sgr~A.
 
The second principal component (Figs. \ref{fig:PCA} and \ref{fig:PCA_vect})
shows the major difference (10.6  percent of the variance)
among the eight lines, which is differences in the ridge of emission in Sgr~B2
and the dense cores in Sgr~A, relative to that in the first component. 
The lines of HCN, HCO$^{+}$, HNC and (to a lesser extent) N$_{2}$H$^{+}$
show a relative lack of emission, which we attribute (see below, subsection 
\ref{subsec:line_intensities_ratios}) to the brightest areas being optically 
thick. The other lines (SiO, HNCO, 
HC$_{3}$N and CH$_{3}$CN) are, relative to the first component, enhanced
at these peaks, which is partly because they are less likely to be optically 
thick, but it is also likely that these molecules are preferentially
found in dense cores.

The third principal component (Figs. \ref{fig:PCA} and \ref{fig:PCA_vect})
shows small differences (2.8  percent of the variance)
in the Sgr~B2 ridge, between most significantly
SiO and HNCO compared to CH$_{3}$CN. The fourth principal component shows small
differences (1.7  percent of the variance), most significantly between HNCO and 
SiO. Although the fourth component describes only a small amount of the variance,
it is striking that it still appears to have physical significance, as the
positive feature corresponds well to the north cloud of Sgr~B2 
\citep{jo+08,jobulo08}, where HNCO is enhanced \citep{mi+98}.

As Sgr B2 is one of the strongest features in the CMZ, and also a 
chemically complex region, some of the most significant structure in 
the PCA components here at is in the Sgr B2 region. For PCA analysis of the
Sgr B2 area with more lines at 3-mm and 7-mm see \citet{jobulo08} and
\citet{jo+11} respectively.

Fig. \ref{fig:PCA_vect} shows the values of the PCA component vectors,
for the first four components, plotted as pairs (for convenience of plotting
multi-dimensional vectors in two dimensions). These vectors describe how the 
normalised integrated images of different lines
are composed of the sum of different scalar amounts (the components PC 1, 
PC 2 etc) of the principal component images (Fig. \ref{fig:PCA}).
All eight lines have distributions
which are made up of positively scaled amounts (PC 1, $\sim$ 0.3) of 
the first principal component, and
further made up of positive or negative amounts (PC 2, etc) of the 
further principal components.

We also note that the results from PCA analysis 
(Figs. \ref{fig:PCA} and \ref{fig:PCA_vect}) are robust to the potential
problems of integrating the emission discussed above in subsection 
\ref{subsec:integ_em}. The PCA components found are almost identical
using integrated emission with the 3-$\sigma$ cutoff here, to that from
the integrated emission of the baselevel-filtered cubes.

\subsection{Line Intensities and Line Ratios}
\label{subsec:line_intensities_ratios}

The previous subsection (\ref{subsec:pca}) on PCA shows there are
significant differences between the morphology shown by the integrated
emission of different lines (at least for the eight strongest lines
used for the PCA). One of the most commonly used analysis tools for
spectral lines is to consider their line ratios, which we do in this
sub-section in order to quantify the behaviour of the principal
emitting species further across the CMZ.

The ratio of the integrated emission of the $^{12}$C to $^{13}$C
isotopologues of HCN, HCO$^{+}$ and HNC lines confirms that the
emission of the $^{12}$C isotopologue is generally optically thick. It
is usually much less than 24, the isotopic ratio for
[$^{12}$C/$^{13}$C] at the Galactic Centre \citep{lape90}, and the
value that would be found for optically thin emission.  This suggests
that the differences in integrated line distribution of HCN, HCO$^{+}$
and HNC, as shown by the second principal component (subsection
\ref{subsec:pca}, Fig.~\ref{fig:PCA}), are related to optical depth
effects.

\begin{figure*}
\includegraphics[angle=-90, width = 17 cm]{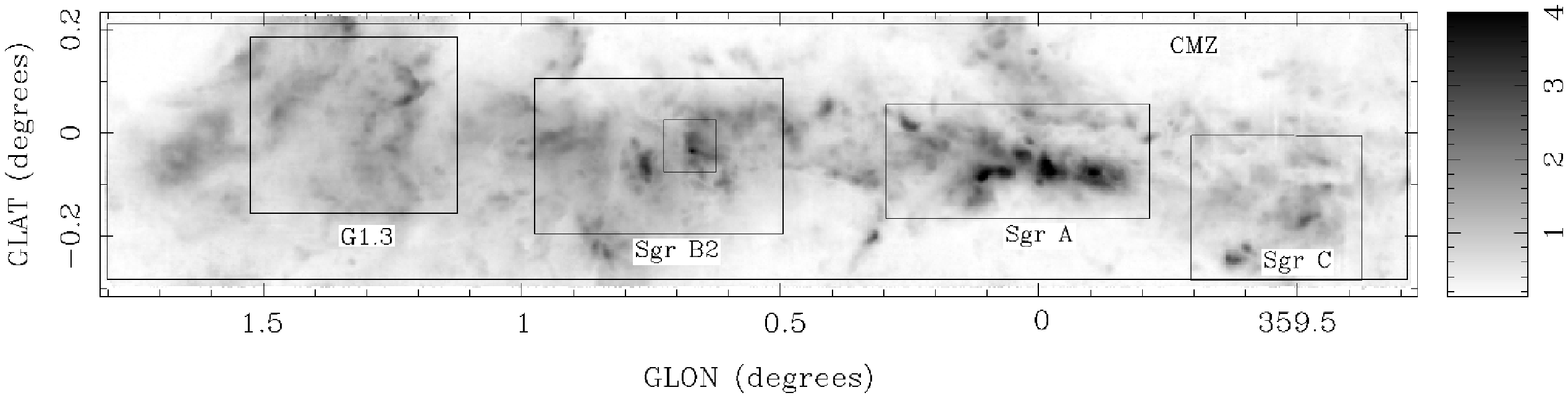}
\caption{The six areas selected for analysis: around Sgr~A, Sgr~B2, Sgr~C, G1.3,
a smaller area at the Sgr~B2 core, and the larger CMZ area, plotted on the peak
HCN emission.}
\label{fig:areas}
\end{figure*}

In order to present an analysis for a manageable part of this
extensive data set we have obtained on the CMZ we have selected
representative regions within the CMZ and examined the line emission
through averaged profiles for these apertures.  The apertures chosen
are listed in Table \ref{tab:apertures}
and plotted in Fig. \ref{fig:areas}. They include the 
entire region of the CMZ we have mapped, and the four main emission `cores'; 
Sgr~A, Sgr~B2, Sgr~C and G1.3.  We also include a smaller region around the 
core of Sgr~B2.  Table \ref{tab:apertures}
lists the centres and sizes of these apertures as well as the velocity
ranges chosen for determining the integrated line emission and the
line ratios from each aperture.  These apertures also provide a basis
for comparing the molecular emission from the centre of the Galaxy to
that from the central regions of other galaxies, where generally the
emission is spatially unresolved and often only the very brightest
molecular lines (i.e.  from HCN, HNC and HCO$^{+}$) can be seen.  We
can compare the equivalent parameters found for the entire CMZ, and
with the distinct emission regions within it, to examine how they
might vary across the central regions of the Galaxy.  We can also
determine the same parameters in the optically thin isotopologue lines
in the CMZ (i.e. H$^{13}$CN, HN$^{13}$C and H$^{13}$CO$^{+}$) to
examine whether the optical depth of particular lines may be
distorting any interpretation being placed upon a given line ratio.

\begin{table*}
\begin{center}
  \caption{Apertures selected for analysis. The areas are rectangles
    in Galactic coordinates, centred on $l$ and $b$, and using the
    velocity ranges indicated to include significant line emission. The 
    projected area on the sky assumes a distance of 8.0 kpc.}
\label{tab:apertures}
\begin{tabular}{ccccccccc}
\hline
Source     &   $l$   & $b$ & Width  & Height & Area         & Area   & Min. vel.
   & Max. vel. \\
           &   deg   & deg & arcmin & arcmin & arcmin$^{2}$ & pc$^2$ & 
km s$^{-1}$ & km s$^{-1}$ \\
\hline
CMZ         &  0.545 & -0.035  & 151.0 & 29.8 & 4\,500 & 24\,400 & -220 & 220 \\
Sgr~A       &  0.040 & -0.055  &  30.8 & 13.4 &    413 &  2\,240 & -170 & 220 \\
Sgr~B2      &  0.735 & -0.045  &  29.0 & 18.2 &    528 &  2\,860 & -130 & 210 \\
Sgr~C       & -0.460 & -0.145  &  20.0 & 17.0 &    340 &  1\,840 & -200 & 170 \\
G1.3        &  1.325 &  0.015  &  24.2 & 20.6 &    499 &  2\,700 &  -90 & 220 \\
Sgr~B2 Core &  0.675 & -0.025  &   6.2 &  6.2 &     38 &    210 &  -20 & 200 \\
\hline
\end{tabular}
\end{center}
\end{table*}

\begin{figure*}
\includegraphics[height=22 cm]{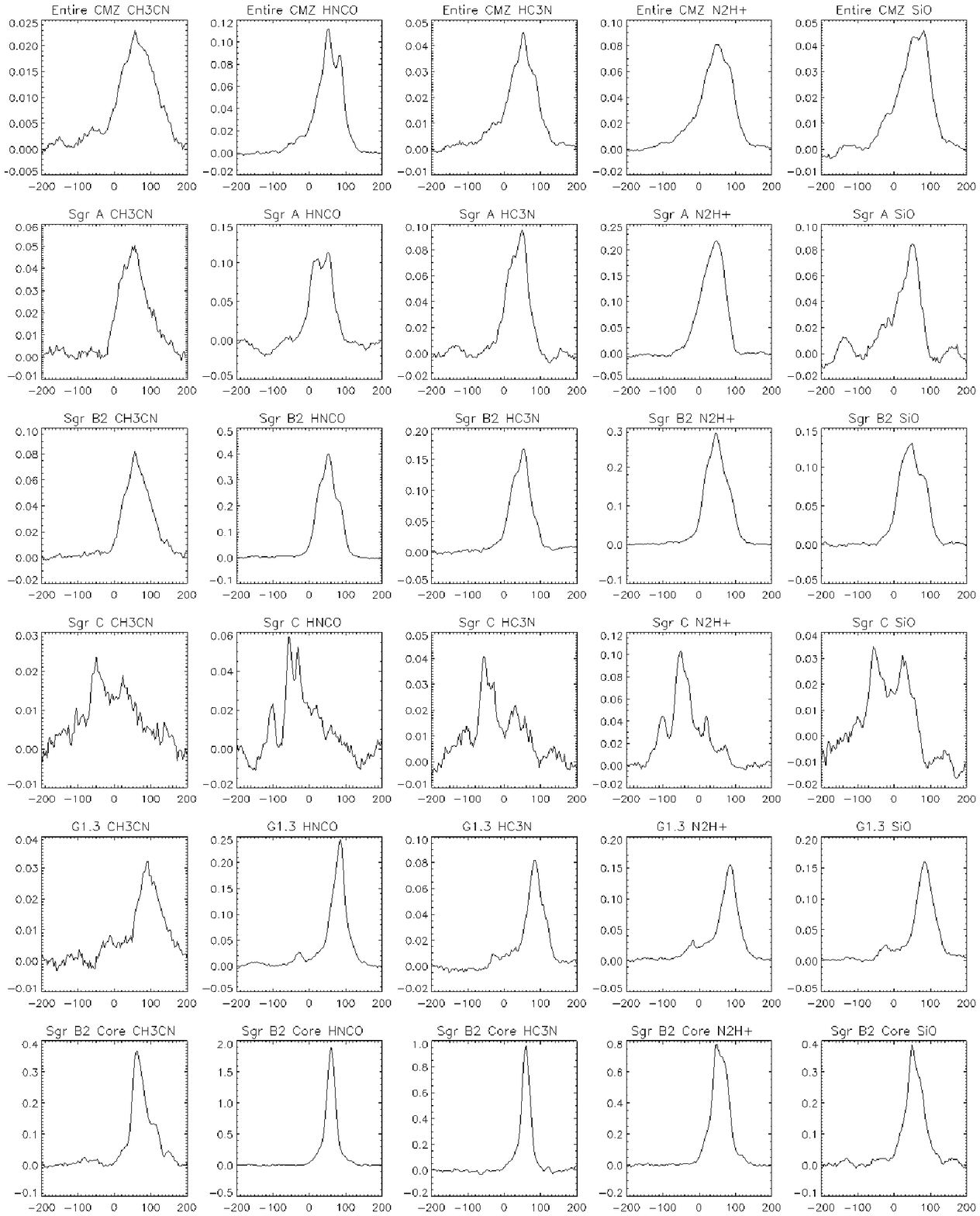}
\caption{Profiles of five molecular lines in six selected apertures in the 
CMZ, as defined in Table \ref{tab:apertures} and Fig. \ref{fig:areas}.
From left to right, the lines are (see Table \ref{tab:lines_table})
from the  CH$_{3}$CN, HNCO, HC$_{3}$N, N$_{2}$H$^{+}$ and SiO molecules.  The 
apertures run from top to bottom and are: the entire CMZ, Sgr~A, Sgr~B2, 
Sgr~C, G1.3 and the Sgr~B2 Core.  Intensities are the averaged pixel values 
in each aperture, in Kelvin, and are $T_{A}^{*}$ (i.e. without beam efficiency 
correction).  The $x$-axis is the velocity, $V_{LSR}$, running from -200 to 
+200 km s$^{-1}$.}
\label{fig:misc_profiles}
\end{figure*}

\begin{figure*}
\includegraphics[height=22 cm]{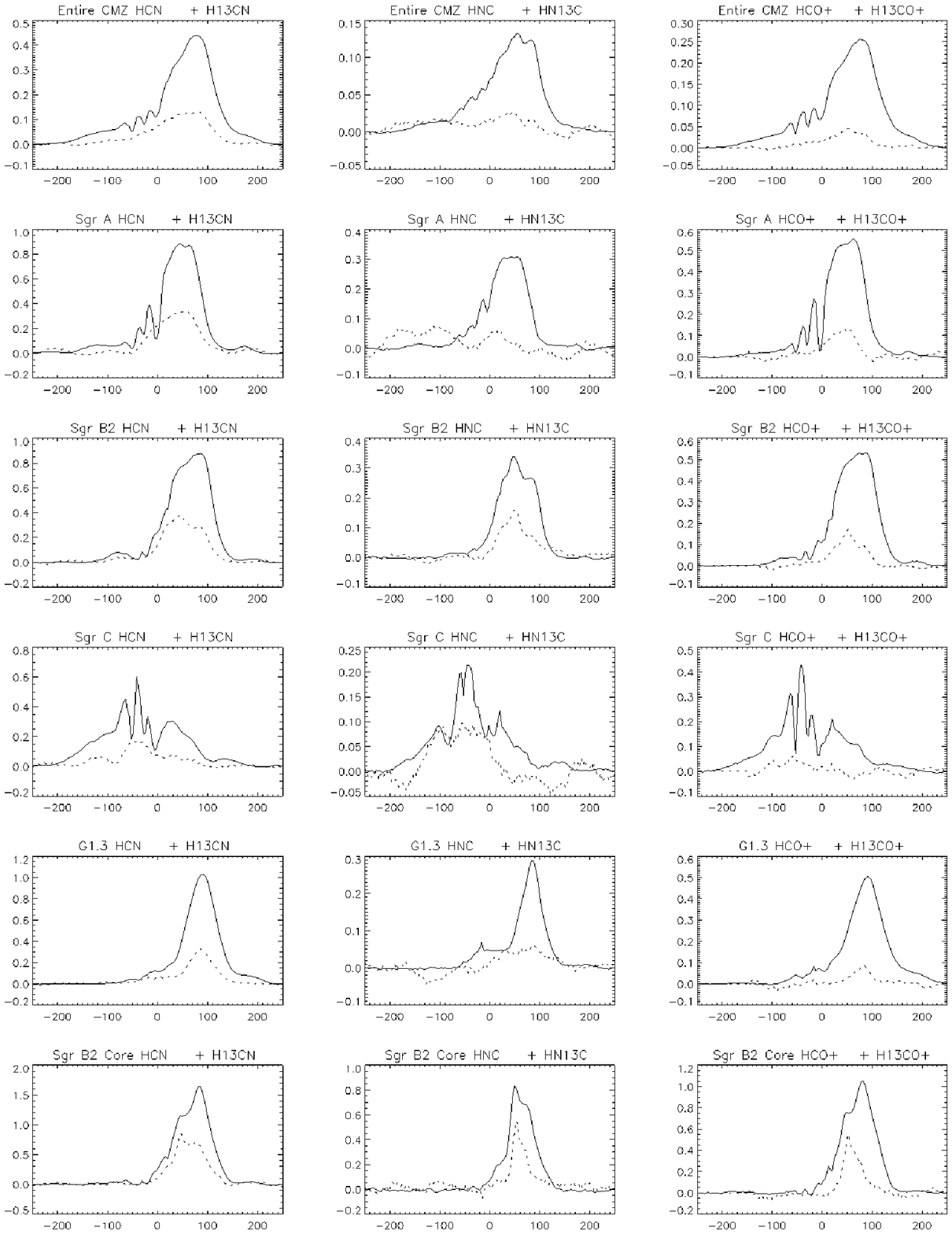}
\caption{Profiles of the three bright molecular transitions, the HCN, HNC 
and HCO$^{+}$ J=1-0 lines in the Central Molecular Zone (left, middle and 
right columns, respectively).  Overplotted as dashed lines are the 
corresponding isotopologues; H$^{13}$CN, HN$^{13}$C and H$^{13}$CO$^{+}$ J=1-0,
multiplied by a factor of 3 for clarity.  
From top to bottom are the six selected apertures 
(see Table \ref{tab:apertures} and Fig. \ref{fig:areas}): the entire 
CMZ, Sgr~A, Sgr~B2, Sgr~C, G1.3 and Sgr~B2 Core. Intensities ($T_{A}^{*}$) 
are the averaged values over each aperture and velocities are $V_{LSR}$ and 
run from -250 to +250 km s$^{-1}$.}
\label{fig:hcn_etc_profiles}
\end{figure*}

In addition to the HCN, HNC and HCO$^{+}$ lines, 
as listed in Table \ref{tab:lines_table}, there 
are other bright lines whose distribution is extended across the CMZ.  We show 
the profiles for the  CH$_{3}$CN, HNCO, HC$_{3}$N, N$_{2}$H$^{+}$ and SiO 
lines in Fig. \ref{fig:misc_profiles},
together with those for HCN, HNC and HCO$^{+}$ (and their isotopologues) in Fig.
\ref{fig:hcn_etc_profiles}.
For each aperture the profiles of all lines are rather similar and are
very broad (over 100 km s$^{-1}$ wide).  Some profiles show several
absorption features, for instance at $-52$, $-28$ and $-3$ km s$^{-1}$
attributable to absorption by cold, foreground gas, likely in spiral
arms along the sight line (e.g.\ as also seen in the AST/RO CO 4--3
datacube of the CMZ; \citet{ma+04}). The brightest of the lines is HCN, but
in some instances other lines are comparably bright (e.g. HNCO in the
Sgr~B2 Core).  The brightness of HCN is generally twice as strong as
HCO$^{+}$, the second brightest of these lines, and three times
stronger than HNC.  Their isotopologue lines are significantly weaker,
and we use the ratios with the main lines to derive optical depths in
subsection~\ref{subsec:opticaldepth}.

Table \ref{tab:mean_fluxes}
presents the integrated line fluxes for these apertures, in K~km~s$^{-1}$ 
and corrected for the extended aperture efficiency of 0.65 \citep{la+05}.  
For comparison with extragalactic line luminosities $L'$, we integrated 
the brightness temperatures spatially and spectrally to obtain values in 
units of K\,km\,s$^{-1}$\,pc$^{2}$ in the different apertures
listed in Table~\ref{tab:fluxarea}. As a reference, we also derived 
$L'_{\rm CO}$ 
using the integrated intensity map of the Columbia survey \citep*{dahath01}. 
Over the entire CMZ, $L'_{\rm CO}$ amounts to 
$2.2\times10^{7}$\,K\,km\,s$^{-1}$\,pc$^{2}$. Relative to CO, the dense 
gas tracers exhibit $L'_{\rm dense}/L'_{\rm CO}$ luminosity ratios of 
0.095 (HCN), 0.010 (H$^{13}$CN), 0.029 (HNC), 0.002 (HN$^{13}$C), 0.057 
(HCO$^{+}$), 0.002 (H$^{13}$CO$^{+}$), 0.005 (CH$_{3}$CN), 0.014 (HNCO), 
0.007 (HC$_{3}$N), 0.014 (N$_{2}$H$^{+}$), and 0.008 (SiO). 

Table \ref{tab:line_ratios_HCN}
then presents the relative line ratios with respect to the (brightest)
HCN line.  The HCN, HNC and HCO$^{+}$ lines may be optically thick,
and so to determine whether this influences the derived line ratios we
also calculate the isotopologue ratios with H$^{13}$CN in Table
\ref{tab:line_sel_ratios}, alongside the main line ratios with HCN. 
While the HNC/HCN ratio is found to be roughly the same as
HN$^{13}$C/H$^{13}$CN ($\sim$0.3) we find that the HCO$^{+}$/HCN ratio
($\sim$0.6) is typically 2-3 times as high as the
H$^{13}$CO$^{+}$/H$^{13}$CN ratio ($\sim$ 0.2-0.3).  As we will see
below in the analysis of the optical depths, this arises because the
HCN line is moderately optically thick (with a similar value to HNC)
while the HCO$^{+}$ line has optical depth around unity.

\begin{table*}
\begin{center}
  \caption{Mean integrated line fluxes (K km s$^{-1}$), as $\int
    T_{MB} dV$ (i.e.  corrected for aperture efficiency) averaged over
    the apertures and integrated over the velocity ranges given in
    Table \ref{tab:apertures}.  For each molecule the first line gives
    the integrated flux and the second line its statistical 1 $\sigma$
    error.  Note that for the weaker lines the uncertainty in where
    to place the baseline across the wide line profile means that the
    formal S/N for the integrated flux is often low, even though the
    lines themselves are clearly detected. The CO fluxes are obtained from 
    \citet{dahath01}.}
\label{tab:mean_fluxes}
\begin{tabular}{ccccccccccccc}
\hline
Source & CO & HCN  & H$^{13}$CN & HNC & HN$^{13}$C & HCO$^{+}$ & H$^{13}$CO$^{+}$ &
CH$_{3}$CN & HNCO & HC$_{3}$N & N$_{2}$H$^{+}$ & SiO \\
\hline 
CMZ         &    914 &  87 &  9 & 26 &  2 &  52 &  2 &  4 &   13 &  6 & 13 &  7 
\\
            &        &   1 &  1 &  1 &  1 &   1 &  1 &  1 &    1 &  1 &  1 &  1 
\\
Sgr~A       &  1\,200 & 140 & 18 & 48 &  4 &  85 &  4 &  8 &   10 & 10 & 25 &  9 
\\
            &        &   1 &  2 &  2 &  4 &   2 &  3 &  2 &    4 &  2 &  2 &  4 
\\ 
Sgr~B2      &  1\,170 & 142 & 19 & 45 &  5 &  86 &  5 & 11 &   40 & 18 & 33 & 17 
\\
            &        &   1 &  1 &  2 &  2 &   1 &  1 &  1 &    2 &  1 &  1 &  2 
\\
Sgr~C       &  1\,060 & 100 &  9 & 35 &  3 &  61 &  1 &  5 &    5 &  6 & 13 &  3 
\\
            &        &   2 &  1 &  2 &  3 &   2 &  3 &  1 &    3 &  2 &  1 &  4 
\\
G1.3        &  1\,270 & 132 & 12 & 30 &  3 &  73 &  2 &  4 &   20 &  8 & 17 & 17 
\\
            &        &   1 &  1 &  1 &  1 &   1 &  2 &  1 &    2 &  1 &  1 &  2 
\\
Sgr~B2 Core &  1\,610 & 199 & 31 & 76 & 11 & 123 & 10 & 33 &  101 & 48 & 68 & 34 
\\
            &        &   3 &  3 &  5 &  3 &   2 &  4 &  3 &    3 &  3 &  2 &  2 
\\
\hline
\end{tabular}
\end{center}
\end{table*}

\begin{table*}
\begin{center}
  \caption{Line luminosities ($\times 10^4$ K km s$^{-1}$ pc$^2$), 
    $\int T_{MB} dV dA$ integrated over
    each aperture and the velocity ranges given in
    Table \ref{tab:apertures}.   For each molecule the first line gives
    the line luminosity  and the second line its statistical 1 $\sigma$
    error. Note that for the some lines the uncertainty in where
    to place the baseline across the wide line profile means that the
    formal S/N for the luminosity is poor, even though the
    lines themselves are clearly detected. The CO luminosities are obtained from
 \citet{dahath01}.}
\label{tab:fluxarea}
\begin{tabular}{ccccccccccccc}
\hline
Source & CO & HCN  & H$^{13}$CN & HNC & HN$^{13}$C & HCO$^{+}$ & H$^{13}$CO$^{+}
$ &
CH$_{3}$CN & HNCO & HC$_{3}$N & N$_{2}$H$^{+}$ & SiO \\
\hline 
CMZ         & 2\,230 &   212 &   22 &   64 &   5 &  128 &   5 &   11 &   32 &   
16 &   32 &   17 \\
            &       &     1 &    1 &    2 &   2 &    1 &   2 &    2 &    2 &    
1 &    2 &    3 \\
Sgr~A       &   268 &  31.4 &  4.1 & 10.8 & 0.8 & 19.1 & 0.9 &  1.7 &  2.3 &  
2.1 &  5.7 &  2.1 \\
            &       &   0.2 &  0.4 &  0.5 & 0.9 &  0.5 & 0.8 &  0.4 &  0.8 &  
0.3 &  0.5 &  0.8 \\
Sgr~B2      &   335 &  40.6 &  5.4 & 12.9 & 1.5 & 24.6 & 1.5 &  3.1 & 11.4 &  
5.2 &  9.3 &  4.8 \\
            &       &   0.3 &  0.3 &  0.5 & 0.5 &  0.3 & 0.3 &  0.3 &  0.5 &  
0.4 &  0.2 &  0.4 \\
Sgr~C       &   195 &  18.4 &  1.7 &  6.4 & 0.6 & 11.3 & 0.2 &  0.9 &  1.0 &  
1.1 &  2.4 &  0.6 \\ 
            &       &   0.3 &  0.2 &  0.3 & 0.5 &  0.4 & 0.5 &  0.3 &  0.6 &  
0.3 &  0.2 &  0.8 \\
G1.3        &   343 &  35.6 &  3.4 &  8.0 & 0.7 & 19.6 & 0.5 &  1.2 &  5.3 &  
2.2 &  4.5 &  4.7 \\
            &       &   0.4 &  0.2 &  0.3 & 0.4 &  0.2 & 0.5 &  0.2 &  0.5 &  
0.3 &  0.3 &  0.5 \\
Sgr~B2 Core &  33.4 &  4.14 & 0.64 & 1.59 & 0.24 & 2.56 & 0.20 & 0.68 & 2.09 & 
1.00 & 1.42 & 0.71 \\
            &       &  0.06 & 0.06 & 0.11 & 0.06 & 0.05 & 0.08 & 0.06 & 0.06 & 
0.06 & 0.05 & 0.05 \\
\hline
\end{tabular}
\end{center}
\end{table*}

\begin{table*}
\begin{center}
  \caption{Ratios of integrated lines fluxes with HCN. For each
    molecule the first line gives the ratio of that line and the HCN
    1-0 flux and the second line the 1~$\sigma$ error.  For HCN itself
    the errors indicate the flux error range for that molecule.}
\label{tab:line_ratios_HCN}
\begin{tabular}{cccccccccccc}
\hline
Source & HCN  & H$^{13}$CN & HNC & HN$^{13}$C & HCO$^{+}$ & H$^{13}$CO$^{+}$ &
CH$_{3}$CN & HNCO & HC$_{3}$N & N$_{2}$H$^{+}$ & SiO \\
\hline 
CMZ         &  1.00 & 0.10 & 0.30 & 0.03 & 0.60 & 0.03 & 0.05 & 0.15 & 0.07 & 0.
15 & 0.08 \\  
            &  0.01 & 0.01 & 0.01 & 0.01 & 0.01 & 0.01 & 0.01 & 0.01 & 0.01 & 0.
01 & 0.02 \\
Sgr~A       &  1.00 & 0.13 & 0.34 & 0.03 & 0.61 & 0.03 & 0.05 & 0.08 & 0.07 & 0.
18 & 0.07 \\
            &  0.01 & 0.01 & 0.02 & 0.03 & 0.02 & 0.02 & 0.01 & 0.03 & 0.01 & 0.
02 & 0.03 \\
Sgr~B2      &  1.00 & 0.13 & 0.32 & 0.04 & 0.61 & 0.04 & 0.08 & 0.28 & 0.13 & 0.
23 & 0.12 \\
            &  0.02 & 0.01 & 0.02 & 0.01 & 0.01 & 0.01 & 0.01 & 0.01 & 0.01 & 0.
01 & 0.01 \\
Sgr~C       &  1.00 & 0.09 & 0.35 & 0.03 & 0.61 & 0.01 & 0.05 & 0.05 & 0.06 & 0.
13 & 0.03 \\
            &  0.04 & 0.01 & 0.02 & 0.03 & 0.03 & 0.03 & 0.01 & 0.03 & 0.02 & 0.
02 & 0.04 \\
G1.3        &  1.00 & 0.09 & 0.22 & 0.02 & 0.55 & 0.01 & 0.03 & 0.15 & 0.06 & 0.
13 & 0.13 \\
            &  0.02 & 0.01 & 0.01 & 0.01 & 0.01 & 0.01 & 0.01 & 0.02 & 0.01 & 0.
01 & 0.02 \\
Sgr~B2 Core &  1.00 & 0.16 & 0.38 & 0.06 & 0.62 & 0.05 & 0.16 & 0.51 & 0.24 & 0.
34 & 0.17 \\
            &  0.03 & 0.02 & 0.03 & 0.02 & 0.02 & 0.02 & 0.02 & 0.02 & 0.02 & 0.
02 & 0.01 \\
\hline
\end{tabular}
\end{center}
\end{table*}

\begin{table*}
\begin{center}
  \caption{Ratios of integrated line fluxes for HNC and HCO$^{+}$ with
    HCN, together with the same ratios measured for their
    isotopologues.}
\label{tab:line_sel_ratios}
\begin{tabular}{ccccc}
\hline
Source & HNC / HCN & HN$^{13}$C / H$^{13}$CN & HCO$^{+}$ / HCN & H$^{13}$CO$^{+}
$ / H$^{13}$CN \\
\hline
CMZ         & 0.30 $\pm$ 0.01 & 0.25 $\pm$ 0.10 & 0.60 $\pm$ 0.01 & 0.25 $\pm$ 0.09 \\
Sgr~A       & 0.34 $\pm$ 0.02 & 0.20 $\pm$ 0.24 & 0.61 $\pm$ 0.02 & 0.23 $\pm$ 0.22 \\ 
Sgr~B2      & 0.32 $\pm$ 0.02 & 0.28 $\pm$ 0.11 & 0.61 $\pm$ 0.01 & 0.27 $\pm$ 0.08 \\ 
Sgr~C       & 0.35 $\pm$ 0.02 & 0.35 $\pm$ 0.34 & 0.61 $\pm$ 0.03 & 0.11 $\pm$ 0.29 \\ 
G1.3        & 0.22 $\pm$ 0.01 & 0.22 $\pm$ 0.12 & 0.55 $\pm$ 0.01 & 0.15 $\pm$ 0.17 \\  
Sgr~B2 Core & 0.38 $\pm$ 0.03 & 0.37 $\pm$ 0.13 & 0.62 $\pm$ 0.02 & 0.31 $\pm$ 0.15 \\ 
\hline
\end{tabular}
\end{center}
\end{table*}

Molecules like CH$_{3}$CN, HNCO and HC$_{3}$N 
and N$_{2}$H$^{+}$ 
are generally found in dense molecular cores or cold dust cores where
massive star formation has recently been initiated.  The ratios of the
lines from these molecules to HCN (Table~\ref{tab:line_ratios_HCN}) is
seen to be more variable between the apertures than they are for HNC
and HCO$^{+}$, with, in particular, Sgr~B2 showing significantly larger
values of the ratios than the other locations (typically 3-5 times
larger).  SiO, which is more sensitive to the presence of shocks,
shows less variation between the apertures, but still has a ratio
twice as large for Sgr~B2 as in Sgr~A and Sgr~C. However G1.3 has
a similar ratio to Sgr~B2, suggesting the influence of shocks in this
region too.

\begin{figure*}
\includegraphics[height=22 cm]{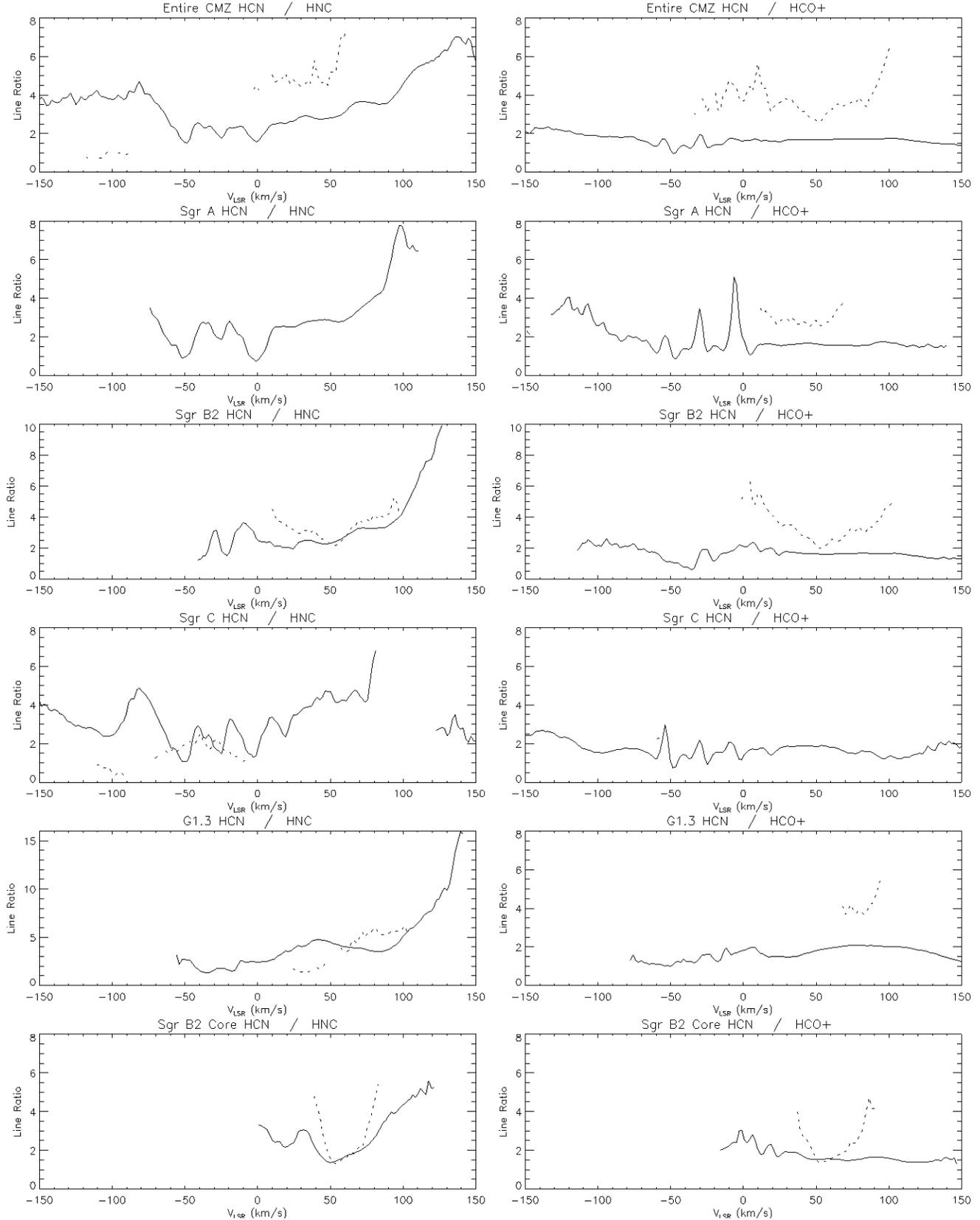}
\caption{Line ratios, as a function of velocity, for HCN/HNC (left) and 
HCN/HCO$^{+}$ (right), for the six selected apertures in Table 
\ref{tab:apertures}
(the entire CMZ, Sgr~A, Sgr~B2, Sgr~C, G1.3 and the Sgr~B2 Core).  The
main line is shown in solid and the $^{13}$C--isotopologue ratios
(H$^{13}$CN/HN$^{13}$C and H$^{13}$CN/H$^{13}$CO$^{+}$) are dashed.
Line ratios are only shown when the intensities of both lines are $>
5~\sigma$ in the relevant velocity channel.  Note that for HN$^{13}$C
in Sgr~A this results in no useable isotopologue data (and just 2 data
points for H$^{13}$CO$^{+}$ in Sgr~C).  The $x$-axis is the $V_{LSR}$
velocity, running from -150 to +150 km s$^{-1}$. Note that the
$y$-axis ranges are the same in all plots except for Sgr~B2 and G1.3
(left column).}
\label{fig:hcn_etc_ratios}
\end{figure*}

Table \ref{tab:max_ratios}
lists the maximum and minimum line ratios found for HCN, HNC and
HCO$^{+}$, as well as for their $^{13}$C isotopologues, as a function of
velocity between the apertures, as opposed to the integrated line
fluxes.  The variations between these extrema can be seen in Fig.
\ref{fig:hcn_etc_ratios}
(note that line ratios are only shown when the individual lines have
S/N~$> 5$ for a given velocity channel).  Within the core of each line
the ratios are generally constant, aside from the velocity ranges
where foreground absorption is evident (see above).  The isotopologue
line ratios are generally similar for H$^{13}$CN / HN$^{13}$C compared
to HCN/HNC. However for H$^{13}$CN / H$^{13}$CO$^{+}$ the
isotopologue line ratio is typically twice the value of HCN/HCO$^{+}$,
as the velocity varies.  This result, for the variation in ratio as a
function of velocity, is also consistent with behaviour found for the
ratio of the integrated fluxes of these lines.

However, at the highest red-shifted velocities for emission 
($> 100$~km s$^{-1}$), the behaviour changes, with the ratio HCN / HNC
increasing significantly, from a value of $\sim 2$ over most of the
profile, to $\sim 10$ at its high-velocity limit.  This trend is also apparent
in the $^{13}$C forms, H$^{13}$CN/HN$^{13}$C, though lower S/N limits
the velocity range over which the comparison may be made. We note,
however, that the HCN / HCO$^{+}$ ratio does not show any evidence for
increasing at these high red-shifted velocities, suggesting that this
variation is a result of the relative strength of the HNC line
decreasing with increasing velocity, while the relative intensity of
the HCN and HCO$^{+}$ lines remains constant.

\begin{table*}
\begin{center}
  \caption{Maximum and minimum values of selected line ratios with the
    velocities ($V_{LSR}$) at which these extrema occur. Note that
    line intensities need to be $> 5 \sigma$ per channel to be used
    for this analysis. No data meets this criteria for HN$^{13}$C in
    Sgr~A.}
\label{tab:max_ratios}
\begin{tabular}{ccccccccccccccccc}
  \hline
  Source &  \multicolumn{4}{c}{HCN / HNC}  &  \multicolumn{4}{c}{H$^{13}$CN / HN
$^{13}$C}  & \multicolumn{4}{c}{HCN / HCO$^{+}$} & 
  \multicolumn{4}{c}{H$^{13}$CN / H$^{13}$CO$^{+}$} \\
  & Max. &  Vel. &  Min. &  Vel. & Max. &  Vel. &  Min. &  Vel. & Max. &  Vel. &
  Min. &  Vel. & Max. &  Vel. &  Min. &  Vel. \\
  & ratio & kms$^{-1}$ & ratio & kms$^{-1}$ & ratio & kms$^{-1}$ & ratio & kms$^{-1}$ & ratio & kms$^{-1}$ & ratio & kms$^{-1}$ & ratio & kms$^{-1}$ & ratio & kms$^{-1}$\\
  \hline
  CMZ         &  7.5 & 183 & 1.5 & -49 & 8.2 &  77 & 0.7 &-116 & 2.5 & -192 & 0.
9 & 223 & 6.6 & 101 & 2.6 & 52 \\
  Sgr~A       &  7.8 &  97 & 0.7 &  -1 & -   &  -  & -   & -   & 5.1 &   -7 & 0.
9 & -47 & 3.8 &  68 & 2.6 & 45 \\ 
  Sgr~B2      &  9.9 & 127 & 1.2 & -41 & 5.3 &  94 & 2.2 &  54 & 2.6 &  -94 & 0.
6 & -36 & 6.3 &   4 & 2.0 & 54 \\ 
  Sgr~C       &  6.8 &  81 & 1.1 & -50 & 2.5 & -39 & 0.4 & -92 & 3.0 &  -54 & 0.
7 & -49 & 2.5 & -56 & 2.3 &-59 \\ 
  G1.3        & 15.9 & 139 & 1.3 & -34 & 6.2 & 101 & 1.4 &  34 & 2.1 &   75 & 1.
0 & 212 & 5.4 &  94 & 3.7 & 70 \\  
  Sgr~B2 Core &  5.6 & 117 & 1.4 &  50 & 5.5 &  86 & 1.3 &  54 & 3.0 &   -1 & 1.
3 & 143 & 4.7 &  86 & 1.4 & 52 \\ 
  \hline
\end{tabular}
\end{center}
\end{table*}

\subsection{Optical Depth Analysis}
\label{subsec:opticaldepth}

\begin{table*}
\begin{center}
  \caption{Maximum and minimum values of the ratios of the line
    intensities, per velocity channel, for HCN, HNC and HCO$^{+}$ with
    their $^{13}$C isotopologues, together with the velocities
    ($V_{LSR}$) at which these extrema occur. Note that all line
    intensities need to be $> 5 \sigma$ per channel to be used for
    this analysis. No data meets this criteria for HN$^{13}$C in
    Sgr~A.}
\label{tab:maxmin_isotopologues}
\begin{tabular}{ccccccccccccc}
\hline
Source &  \multicolumn{4}{c}{HCN / H$^{13}$CN}  &  \multicolumn{4}{c}{HNC / HN$^{13}$C}  
& \multicolumn{4}{c}{HCO$^{+}$ / H$^{13}$CO$^{+}$} \\
& Max. &  Vel. &  Min. &  Vel. & Max. &  Vel. &  Min. &  Vel. & Max. &  Vel. &  
Min. &  Vel. \\
& ratio & km~s$^{-1}$ & ratio & km~s$^{-1}$ & ratio & km~s$^{-1}$ & ratio & km~s$^{-1}$ & ratio & km~s$^{-1}$ & ratio & km~s$^{-1}$  \\
\hline
CMZ         & 22.3 & 139 & 4.6 & -1   & 23.3 &  77 & 2.1 & -114 & 44.1 & 101 & 1
0.1 & -28 \\
Sgr~A       & 11.1 & -37 & 1.5 & -1   & -    &  -  & -   & -    & 20.3 &  68 & 1
2.4 &  46 \\ 
Sgr~B2      & 15.8 & 127 & 2.6 & -47  & 14.1 &  94 & 6.2 &   52 & 38.6 & 103 &  
8.4 &  52 \\ 
Sgr~C       & 29.9 & -78 & 2.9 & -54  &  8.9 & -39 & 2.2 &  -83 & 14.2 & -59 &  
7.0 & -56 \\ 
G1.3        & 18.9 & 143 & 3.3 & -36  & 16.4 &  81 & 3.1 &   26 & 27.4 &  94 & 1
6.6 &  83 \\  
Sgr~B2 Core & 10.4 & 110 & 4.1 &  46  & 13.9 &  83 & 4.4 &   54 & 22.3 &  86 &  
4.0 &  52 \\ 
\hline
\end{tabular}
\end{center}
\end{table*}

\begin{figure*}
\includegraphics[height=21 cm]{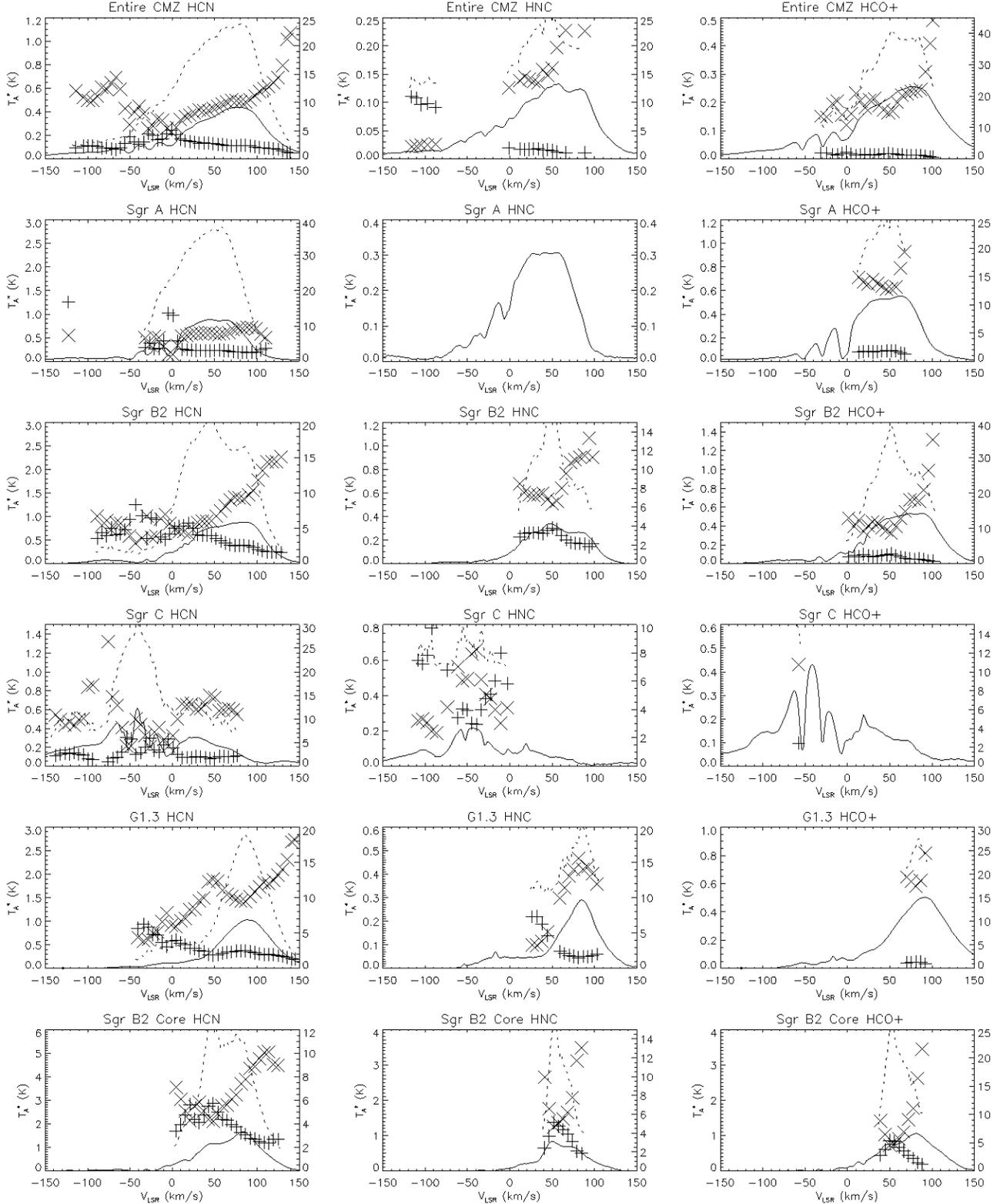}
\caption{Optical depth-corrected line intensities, together with the optical 
depth and isotopologue ratios, as a function of velocity.  From left to right 
are the HCN, HNC and HCO$^{+}$ measured line profiles (solid line) and 
optically-depth corrected profile (dashed line).  The intensity, $T_{A}^{*}$, 
in Kelvin is shown on the left-hand axis, and the $x$-axis shows the velocity in
$V_{LSR}$ from $-150$ to $+150$ km s$^{-1}$.  The symbol $\times$ shows the 
measured isotopologue line ratio for each molecule (i.e. HCN/H$^{13}$CN, 
HNC/HN$^{13}$C and 
HCO$^{+}$/H$^{13}$CO$^{+}$) and the  $+$ sign shows the calculated optical depth 
for the HCN, HNC and HCO$^{+}$ J=1-0 lines, respectively (see text).  The 
right-hand axis indicates the numerical value for the line ratio and optical 
depth. The six selected apertures (see Table \ref{tab:apertures})
are, from top to bottom, the entire CMZ, Sgr~A, Sgr~B2, Sgr~C, G1.3
and the Sgr~B2 Core.  Note that optical depths and line ratios are
only calculated when the corresponding lines are $> 5~\sigma$ (i.e.
HCN and H$^{13}$CN $> 5~\sigma$, etc.).  For some lines/sources (e.g.
HN$^{13}$C in Sgr~A) there are few, or no, data for which this
condition is met. Also, for clarity, every three data points in the
$x$-direction are binned together to provide a single point on the
plots.}
\label{fig:hcn_etc_isotopes}
\end{figure*}

Where the H$^{13}$CN, HN$^{13}$C and H$^{13}$CO$^{+}$ lines are well
detected (S/N~$> 5$) we may use the ratio with the main isotopologue
to determine the optical depth of the emission, as a function of
velocity across the line ratio.  We list in
Table~\ref{tab:maxmin_isotopologues} the maximum and minimum values of
the ratios found for the isotopologue pairs (i.e.\ HCN/H$^{13}$CN,
HNC/HN$^{13}$C and HCO$^{+}$/H$^{13}$CO$^{+}$) across their line profiles,
together with the velocities at which these extrema occur.  Again,
these numbers are filtered to require the S/N in each velocity channel
to be $> 5 \sigma$. The variation of each line ratio with velocity can
be followed in Fig.~\ref{fig:hcn_etc_isotopes}.  The ratios are
generally well less than 24, the abundance ratio for $^{12}$C to
$^{13}$C determined for the centre of the Galaxy \citep{lape90},
indicating that the $^{12}$C isotopologues of these species must be
optically thick.  However, the isotopologue ratios are also
significantly greater than unity, which also implies that the $^{13}$C
lines cannot be strongly optically thick.  We make use of this
behaviour to provide an approximate solution of the radiation transfer
equation below.  We apply a standard analysis technique, which we
summarise here.

The radiative transfer equation has the general solution for the intensity 
of a velocity channel in a spectral line known as the detection equation
\citep[e.g.][]{stpa05}: 
\[ T_{A} = T_{A}^{*} / \eta = f [J_{\nu}(T_{ex}) - J_{\nu}(T_{BG})] (1 - \exp(-\tau_{\nu})) \]
where $T_{A}^{*}$ is the measured intensity, $\eta$ the extended beam 
efficiency (taken as 0.65, \citealp{la+05}).  $f$ is the beam filling factor 
for the emission, 
$ J_{\nu}(T) = [h \nu / k ] / [\exp(h \nu / k T) - 1] $,
with $T_{ex}$ being the excitation temperature and $T_{BG}$ being the 
temperature of the CMBR (i.e. 2.726~K).  $\tau_{\nu}$ is the optical depth of 
the emission.

For a line isotopologue pair, say HCN and H$^{13}$CN, we assume the
former is optically thick, but the latter is optically
thin, as inferred above.
We also assume they have the same excitation temperature and beam
filling factor.  Furthermore,
\[ \tau_{HCN} / \tau_{H^{13}CN} = N(HCN) / N(H^{13}CN) = X_{^{12}C/^{13}C} \] 
the abundance ratio of the $^{12}$C isotope to $^{13}$C, which we have taken 
to be 24 \citep{lape90}, assuming no isotope fractionation in the molecules.

Then we see, from application of the detection equation to HCN and H$^{13}$CN, 
that $\tau_{H^{13}CN} = T_{A}^{*}(H^{13}CN) / T_{A}^{*}(HCN)$, 
with $\tau_{HCN}$ then determined by the preceding formula.

We hence apply this analysis to determine the optical depth of the HCN, HNC 
and HCO$^{+}$ lines, as a function of velocity.  Furthermore, from standard 
molecular radiative transfer theory \citep[e.g.][]{gola99} we may show that the 
column density of the upper level of each transition is given by
\[ N_{u} = [8 \pi k / A h] [\nu^{2} / c^{3}] T_{A} [\tau / 1 -
\exp(-\tau)] \delta V \] where $k$, $h$ and $c$ are the well-known
physical constants, $\nu$ is the frequency of the transition, $A$ its
radiative decay rate (Einstein coefficient) and $\delta V$ the channel 
velocity spacing.  The
optically thin case ($\tau \ll 1$) simply has the optical depth
correction factor $\tau / (1 - \exp(-\tau))$ set to unity.

Once $N_{u}$ has been determined the total molecular column density
$N$ can be found by applying $ N = (N_{u} Q(T)/g_{u}) \exp(E/kT_{ex})$
with $g_{u} = 2J+1 = 3$, the partition function $Q(T) = 2 k T / h
\nu$, $E$ the energy of the upper level and an assumed excitation
temperature, $T_{ex}$.  We take the later as 24~K, based on the dust
temperature of the Sgr~B2 envelope \citep{jo+11}.  \citet{na+09}
determined that $T_{ex}$ has the range 20 to 35~K in the CMZ from LVG
modelling of the CO lines. To check the sensitivity of this assumption
we also repeated our calculations by halving and doubling the
temperature.  For the former (i.e. $T = 12$~K) the column densities (and
masses) decrease by a factor $\sim 0.6$. For the latter (i.e.
$T = 48$~K) they increase by a factor $\sim 1.8$. While undoubtedly
variations in the excitation temperature across the CMZ could result
in relative variations in these calculated quantities by factors of a
few, the absolute determinations for the column density in each
aperture are not likely to be in error by more than a factor 2 due to this
assumption of a constant temperature $T = 24$~K

Similarly the molecule mass $M$ can then be determined from the column
density $N$ knowing the aperture size (Table \ref{tab:apertures}),
molecular masses $m_{mol}$, and the source distance ($D = 8.0$~kpc) as
$ M = N m_{mol} \Omega D^{2}$.

In Table \ref{tab:tau}
we list the derived optical depths for each aperture for HCN, HNC and
HCO$^{+}$.  We list both the maximum value of the optical depth found
at each position (together with the relevant emission velocity) and
the weighted mean optical depth.  This is determined from the value of
the correction factor $\tau / (1 - \exp(-\tau))$ needed to convert the
integrated line flux from that for an optically thin line to the
measured integrated line flux after applying the optical depth
correction at each velocity channel.  This number thus provides a
representative value of the optical depth for each line.  For HCN and
HNC this is typically 2-3, whereas for HCO$^{+}$ it is around unity.
On the other hand, the maximum values of the optical depth for each
line are several times their mean values.  These are illustrated
graphically in Fig.~\ref{fig:hcn_etc_isotopes}.
This shows, for each source and line, both the measured and the
optical-depth corrected line profile.  Also shown, with the right-axis
as the numerical value, are the measured value of the isotopologue
line ratio (i.e. HCN/H$^{13}$CN etc.) and the value determined for the
optical depth at each velocity.  The optical depth is found to be
fairly constant over each line though note that it is higher where the
absorption features are seen; however by applying the optical depth
correction these features are removed from the line profiles, as can
be seen by their smoothly passing across the affected velocities in
Fig.~\ref{fig:hcn_etc_isotopes}.

\begin{table*}
\begin{center}
  \caption{Optical depths, for HCN, HNC, and HCO$^{+}$, derived from
    the isotopologue ratios assuming [$^{12}$C/$^{13}$C] = 24, given
    as the maximum and the weighted mean. The velocity listed is that
    at which the maximum optical depth occurs.  See the text for the
    definition of the weighted mean value of the optical depth.}
\label{tab:tau}
\begin{tabular}{cccccccccc}
\hline
Source  & HCN & HCN & HCN & HNC & HNC & HNC & HCO$^{+}$ & HCO$^{+}$ & HCO$^{+}$ 
\\
        & Max.    & Vel. & Mean  & Max.    & Vel. & Mean  & Max.   & Vel.  & Mea
n \\
        & $\tau$ & km s$^{-1}$ & $\tau$ & $\tau$ & km s$^{-1}$ & $\tau$ & $\tau$
 & km s$^{-1}$ & $\tau$ \\
\hline
CMZ         &  5.2  &  -1.1 &  2.4  &  11.7  & -114.1 & 1.0  &  2.4  & -28.4 &  
0.9 \\
Sgr~A       & 16.5  &  -1.1 &  3.0  &   -    &  -     & -    &  1.9  &  46.3 &  
1.0 \\
Sgr~B2      &  9.1  &  -46.6 & 3.2  &   3.9  &   51.8 & 2.5  &  2.9  &  51.8 &  
1.4 \\
Sgr~C       &  8.2  &  -52.1 & 2.2  &  11.0  &  -83.1 & 2.9  &  3.4  & -55.8 &  
0.1 \\
G1.3        &  7.4  &  -35.7 & 2.2  &   7.6  &   26.3 & 1.6  &  1.4  &  82.8 &  
0.4 \\
Sgr~B2 Core &  5.9  &   46.3 & 3.6  &   5.5  &   53.6 & 2.6  &  5.9  &  51.8 &  
2.0 \\
\hline
\end{tabular}
\end{center}
\end{table*}

While the isotopologue line ratios are not found to vary much over the
cores or blue-wings of each line, this is not the case for the
red-shifted gas, however.  For several apertures (especially the
entire CMZ, Sgr~B2 and G1.3; see Fig.~\ref{fig:hcn_etc_isotopes})
these ratios rise steeply for velocities $> +50$~km s$^{-1}$, from
values of $\sim 5$ to reach 20--40 for the most red-shifted emission
(see also Table~\ref{tab:maxmin_isotopologues}; note that this
analysis has only been applied to data with S/N~$> 5$ per channel so
this is not an artifact of low S/N at the profile edges).  This is as
expected if the optical depth is falling with increasing velocity,
with the line ratio tending towards the optically thin value given by
the [$^{12}$C/$^{13}$C] isotope ratio.  However, we note that the
ratios determined for the entire CMZ and Sgr~B2 for the HCO$^{+}$
isotopologues at the highest velocities are $> 30$, greater than the
assumed isotope abundance ratio of 24 for the CMZ.  We discuss this
further in subsection~\ref{subsec:isotopologue}.

Finally Table \ref{tab:col_den+mass}
presents the derived column density and masses for each aperture and
molecule.  Presented here are the upper level column density, without
any optical depth correction, and the total molecule column density,
correcting for optical depth and assuming an excitation temperature of
24 K.  The molecule masses are then derived for each aperture from the
total column density for that molecule.  The column densities for each
molecule vary little with aperture, though of course the masses do as
they are summed over the aperture.  Around 50\,M$_{\odot}$ of
[HCN+HNC+HCO$^{+}$] exists in the CMZ, 60 percent of this being HCN and 20 percent
each HNC and HCO$^{+}$.  Furthermore, 10-15 percent of the total mass of the
CMZ is associated with each of the regions Sgr~A, Sgr~B2 and G1.3.

We end this section by noting that we have considered calculating the
molecule abundances by making use of the sub-millimetre dust data to
provide total column densities in each aperture, and then ratioing
these values with the relevant molecular column densities.  Three such
data sets have been published, at 1.1\,mm \citep[BOLOCAM,][]{ba+10}, at
450 and 850$\mu$m \citep[SCUBA,][]{pi+00} and at 870$\mu$m 
\citep[LABOCA,][]{sc+09}.  However close examination of these data sets reveals
issues with the determination of the baseline in an extended emission
region such as the CMZ, given the data acquisition methods applied.
Significant levels of low-level emission are missing in the final
images, so yielding incorrect dust column determinations, even before
other uncertainties such as dust temperature and opacities are
considered.  We have therefore not attempted to determine molecular
abundances from our data set, which await further work on the analysis
of the extended dust emission across the CMZ before reliable values
can be calculated.

\begin{table*}
\begin{center}
  \caption{Column density and mass for HCN, HNC and HCO$^{+}$. The
    column densities are the mean over each aperture, and calculated
    as $N_{J=1}$, in the $J=1$ level without the correction for
    optical depth and, as $N$ the sum over all levels, with the
    correction for optical depth ($\tau/(1-\exp(-\tau))$), applied per
    velocity channel and assuming $T_{ex} = 24$~K. The mass, $M$, is
    the sum over the aperture defined in Table \ref{tab:apertures},
    calculated including the correction for optical depth. Note if
    $T_{ex}$ is halved $N$ and $M$ decrease by 0.6 times, while if
    $T_{ex}$ is doubled they increase by 1.8 times.}
\label{tab:col_den+mass}
\begin{tabular}{cccccccccc}
  \hline
  Source      & $N_{J=1}$ & $N$ & $M$ & $N_{J=1}$ & $N$ & $M$  & $N_{J=1}$ & $N$
 & $M$ \\
  & HCN & HCN & HCN & HNC & HNC & HNC & HCO$^{+}$ & HCO$^{+}$ & HCO$^{+}$ \\
  & no corr.  & corr.    & corr. & no corr.  & corr.    & corr.
  & no corr.  & corr.    & corr. \\
  & 10$^{14}$ cm$^{-2}$ & 10$^{14}$ cm$^{-2}$ & M$_{\sun}$ 
  & 10$^{14}$ cm$^{-2}$ & 10$^{14}$ cm$^{-2}$ & M$_{\sun}$ 
  & 10$^{14}$ cm$^{-2}$ & 10$^{14}$ cm$^{-2}$ & M$_{\sun}$ \\
  \hline
  CMZ         & 1.5  & 17  & 29   & 1.0  &  7.0  & 11   & 0.74  & 5.0  & 8.8 \\
  Sgr~A       & 1.6  & 23  & 3.5  & 0.96 &  4.2  & 0.64 & 0.72  & 5.1  & 0.81 \\
  Sgr~B2      & 1.7  & 25  & 4.8  & 0.84 &  9.9  & 1.9  & 0.74  & 6.3  & 1.3 \\
  Sgr~C       & 1.4  & 15  & 1.9  & 0.90 &  12   & 1.5  & 0.60  & 2.8  & 0.37 \\
  G1.3        & 1.4  & 16  & 2.9  & 0.61 &  5.4  & 0.98 & 0.65  & 3.5  & 0.68 \\
  Sgr~B2 Core & 1.8  & 31  & 0.43 & 0.84 &  11   & 0.15 & 0.76  & 7.9  & 0.12 \\
  \hline
\end{tabular}
\end{center}
\end{table*}

\section{Discussion}
\label{sec:disc}

\subsection{CMZ Line Luminosities Compared to External Galaxies}
\label{sec:discuss-}

\citet{gaso04} surveyed 52 
galaxies in HCN(1-0) that were selected to cover a large range of FIR 
luminosities. Their sample galaxies show $L'_{\rm CO}$ values in the 
$5\times10^{8}$ to $1.5\times10^{10}$\,K\,km\,s$^{-1}$\,pc$^{2}$ range. 
This is considerably more than what is measured in the CMZ. Even 
galaxies that are considered Milky Way equivalents, such as IC\,342, 
show over an order of magnitude larger $L'_{\rm CO}$ luminosities. The 
reason for the difference is likely the edge-on orientation of the CMZ, 
the sample selection that picks galaxies with rather large FIR fluxes, 
as well as the fact that the physical size of their single dish beam 
projects to kpc size scales at the distances of the sample galaxies -- 
much larger than the $\sim$350~pc of our CMZ image. The ratio of 
$L'_{\rm HCN}/L'_{\rm CO}$ in the \citet{gaso04} sample, however, is in the 
range 0.03 - 0.18, which is very similar to what we find in the CMZ (0.095). 
This is also in agreement with the more recent observation of \citet{kr+08}. 
They observed a sample of 12 nearby, starburst and active galaxies and 
derive $L'_{\rm HCN}/L'_{\rm CO}$ luminosity ratios of 0.1-0.5 for the 
bulk of their sample of 12 galaxies. The CMZ is at the lower end of that 
range and may indicate that the star formation activity correlates with 
the availability of dense gas relative to CO, a trend established by 
\citet*{sodora92} and \citet{gaso04} and interpreted, e.g., by \citet*{krmctu09}. 
\citet{kr+08} also measure HCO$^{+}$ and they derive $L'_{\rm 
HCO^{+}}/L'_{\rm HCN}$ ratio in the 0.5 to 1.6 range. Our value for the 
CMZ is with 0.6, at the lower end of that distribution.

\subsection{Interpreting Line Ratio Variations for HCN, HNC and HCO$^{+}$
  in External Galaxies}
\label{sec:discuss-ratio}

The ratios of the fluxes of the HCO$^{+}$, HCN and HNC lines are often
measured in studies of extra-galactic sources, where they may provide
the only readily accessible probe of the dense molecular environment
across the entire central regions of a Galaxy  \citep[e.g.][]{lo+08}.
In particular, the ratios of these lines have been used to distinguish
X-ray-dominated regions (XDRs) from photodissociation regions (PDRs)
\citep*{mespis07}, and applied to starburst galaxies \citep{ba+08}.

The spatially resolved ratios for these lines that we have determined
from the CMZ are therefore useful in providing a Galactic analogue to
compare such results with. Within the CMZ it is possible to study how
the ratio may change with position, due to changing excitation
conditions, as well as to see how it appears when blended together in a
single pixel as in an external galactic nucleus.  We may also examine
the effects of opacity on the diagnostic line ratios by comparing them
to the ratio obtained using their optically thin isotopologues.  The
isotopologues may not be measurable in an external galaxy due to their
weakness, yet excitation conditions may be also mis-diagnosed if the
measured line ratios do not reflect the intrinsic line ratio due to
one line of a pair being more affected by optical depth than the
other.

In such analyses of external galaxies generally the logarithm of the
line ratios (e.g.  log[I(HCO$^{+}$)/I(HCN)] etc.) is presented. To aid
this comparison we have therefore re-calculated the line ratios in
Table~\ref{tab:line_sel_ratios} into their logarithm values and
present these in Table~\ref{tab:log_lineratios}, together with the
errors.  For completeness we have also included HCO$^{+}$/HNC in this
table although it can be calculated from the other two ratios. The
value for the $^{13}$C line ratios in
Table~\ref{tab:log_lineratios} can be interpreted as providing the
optically thin ratio for the corresponding $^{12}$C line ratios across
the CMZ.

For log(HNC/HCN) the line ratio varies from $-0.4$ to $-0.6$ between
the apertures.  The $^{13}$C line ratios are consistent with those for
the corresponding $^{12}$C value, within the errors of measurement.
However, this is not the case for log(HCO$^{+}$/HCN). This varies little
between the 6 apertures we have chosen for the $^{12}$C line ratios,
from $-0.21$ to $-0.26$. Yet for the $^{13}$C isotopologues the
corresponding value is typically a factor 2 higher (i.e. ranging from
$-0.5$ to $-0.6$).  For log(HCO$^{+}$/HNC) there is a factor of 2
variation measured between the apertures, from 0.2 to 0.4.  However,
while the measured $^{13}$C isotopologue ratios for these species may
appear to differ from this in Table~\ref{tab:log_lineratios}, the
relevant lines are too weak to discern any significant difference from
the corresponding $^{12}$C ratios.  To summarise, the empirical result
from the CMZ data presented here is that the HCO$^{+}$/HCN ratio needs
to be determined from measurements of the $^{13}$C isotopologues, but
the HNC/HCN ratio can be determined using the main $^{12}$C lines.  We
cannot determine whether the $^{13}$C isotopologues are necessary to
measure the true HCO$^{+}$/HNC ratios.  However the analysis given
earlier, showing that the HCO$^{+}$ optical depth is generally less than
that for HNC, suggests that this is likely to be the case.

\citet{lo+08} examine the HCN, HNC and HCO$^{+}$ line ratios in a number
of IR-luminous galaxies and present plots of their variation for the
three $^{12}$C-species ratios considered in
Table~\ref{tab:log_lineratios}. They discriminate in these plots
between excitation regimes dominated by PDRs from XDRs; HNC and
HCO$^{+}$ are found to be relatively stronger, compared to HCN, in XDRs
than PDRs.  They find that nearly all the sources they studied lie
clearly in the PDR-regime; indeed they need to invoke an additional
heating source (``mechanical heating'') even to explain the ratios
under the PDR regime.  

Applying the results from our CMZ study actually exacerbates this
situation.  All points for the CMZ also lie clearly in the PDR regime
in these plots \citep[see Fig.~1 of][]{lo+08}, which is not unexpected
given the $\rm 10^7 L_{\odot}$ IR luminosity of the CMZ is dominated
by re-processed dust emission. The CMZ data points lie in the same
region of the phase space in these diagrams as the majority of the
galaxies in the study. However, applying the H$^{13}$CO$^{+}$/H$^{13}$CN
ratio as a diagnostic, instead of HCO$^{+}$/HCN, moves the CMZ points
even further from the dividing line between the PDR and XDR regimes.
By analogy, this suggests that this is also likely to be the case for
the majority of galaxies where these diagnostic ratios have been
measured.

\begin{table*}
\begin{center}
  \caption{Logarithm of line ratios, for the integrated flux through
    each aperture, for HCN, HNC and HCO$^{+}$, together with same ratios
    measured for the $^{13}$C isotopologue. Statistical errors are
    also listed for each ratio.}
\label{tab:log_lineratios}
\begin{tabular}{ccccccc}
\hline
Source      & log(HCO$^{+}$/HCN)&	log(H$^{13}$CO$^{+}$/H$^{13}$CN) & log(HNC/HCN)& 
log(HN$^{13}$C/H$^{13}$CN) & log(HCO$^{+}$/HNC)& log(H$^{13}$CO$^{+}$/HN$^{13}$C) \\
\hline
CMZ	    & $-0.22 \pm 0.01$ & $-0.59 \pm 0.13$ & $-0.52 \pm 0.02$ & $-0.61 
\pm 0.16$ & $+0.30 \pm 0.02$ & $+0.02 \pm 0.21$ \\
Sgr~A	    & $-0.22 \pm 0.02$ & $-0.64 \pm 0.29$ & $-0.46 \pm 0.03$ & $-0.70 
\pm 0.35$ & $+0.25 \pm 0.04$ & $+0.06 \pm 0.47$ \\
Sgr~B2	    & $-0.22 \pm 0.01$ & $-0.57 \pm 0.11$ & $-0.50 \pm 0.03$ & $-0.54 
\pm 0.14$ & $+0.28 \pm 0.03$ & $-0.02 \pm 0.20$ \\
Sgr~C	    & $-0.21 \pm 0.04$ & $-0.97 \pm 0.57$ & $-0.46 \pm 0.04$ & $-0.45 
\pm 0.30$ & $+0.25 \pm 0.05$ & $-0.52 \pm 0.65$ \\
G1.3	    & $-0.26 \pm 0.02$ & $-0.83 \pm 0.33$ & $-0.65 \pm 0.03$ & $-0.67 
\pm 0.21$ & $+0.39 \pm 0.03$ & $-0.15 \pm 0.42$ \\
Sgr~B2 Core & $-0.21 \pm 0.03$ & $-0.50 \pm 0.18$ & $-0.42 \pm 0.05$ & $-0.43 
\pm 0.14$ & $+0.21 \pm 0.05$ & $-0.07 \pm 0.22$ \\
\hline
\end{tabular}
\end{center}
\end{table*}

\subsection{Isotopologue Line Ratio Variations with Velocity}
\label{subsec:isotopologue}
As can be seen in Fig.~\ref{fig:hcn_etc_isotopes} and was discussed in
subsection~\ref{subsec:opticaldepth} the line ratios for the three
isotopologue pairs increase significantly with red-shifted velocity in
several locations (CMZ, Sgr~B2, G1.3). The trend is particularly
noticeable for HCO$^{+}$ and HCN, and less so for HNC.  This can be
caused by decreasing optical depth of the $^{12}$C species with
increasing velocity, with the line ratio then tending towards the
[$^{12}$C/$^{13}$C] isotope ratio (at least in the absence of
fractionation effects).  However we note that there is no trend of
increasing ratio (or decreasing optical depth) seen to the
blue-shifted side of the line profiles.

Of the three line ratio pairs, the HCO$^{+}$/H$^{13}$CO$^{+}$ ratio is
generally the highest.  This is also consistent with the average
optical depth of the HCO$^{+}$ line being lower than HCN or HNC (see
Table~\ref{tab:tau}). HCO$^{+}$ provides, therefore, the best probe of
the optically thin ratio; i.e.\ of the [$^{12}$C/$^{13}$C] isotope
ratio.

For the Sgr~B2 and integrated CMZ apertures the HCO$^{+}$/H$^{13}$CO$^{+}$
ratio rises as high as $\sim 40$ at the high-velocity edge of the profile,
considerably in excess of the [$^{12}$C/$^{13}$C] average isotope
ratio of 24 for the centre of the Galaxy.  Taken by itself, the rising
ratio with decreasing HCO$^{+}$ intensity in the red-shifted wing could be a
result of a systematic error in the baseline subtraction of the
$^{13}$C species, making the calculated flux relatively fainter
compared to the $^{12}$C species than it should be.  However a similar
result has also been reported by \citet{ri+10b} in the Galactic Centre
from observations using the IRAM 30m telescope.  They find similarly
high HCO$^{+}$/H$^{13}$CO$^{+}$ ratios ($\sim 40$) at several selected
positions in the central $5^{\circ}$ of the Galaxy, with the ratio
also highest for the most red-shifted gas.  \citet{ri+10b} also
measure lower values for the HCN and HNC isotopologue ratios, similar
to our results.

If these inferred isotope ratios of $> 24$ are correct then this
red-shifted gas has undergone a different degree of nuclear processing
that the bulk of the CMZ gas, which has a [$^{12}$C/$^{13}$C] isotope
ratio of 24.  $^{12}$C is primarily produced in first generation,
massive stars on a rapid timescale. $^{13}$C arises from CNO
processing of $^{12}$C in lower mass stars, on a slower time-scale
\citep[e.g.][]{pr+96}. There is also a gradient in [$^{12}$C/$^{13}$C]
observed with distance from the centre of the Galaxy, rising from
$\sim 24$ there to reach 80-90 in the solar neighbourhood
\citep{wi99}. \citet{ri+10b} interpret these elevated isotope ratios
resulting from infall of material to the centre of the Galaxy.
Positions observed associated with $x_{1}$ orbits (which are parallel to
the Bar) had higher inferred values of [$^{12}$C/$^{13}$C] than those
positions associated with $x_{2}$ orbits (which are inside and orthogonal
to the $x_{1}$ orbits; see \citet{st+04}).

There is limited overlap between the positions observed by
\citet{ri+10b} and ourselves; moreover the large apertures we have
used for our analysis are in contrast to the single beams used by
\citet{ri+10b}.  Nevertheless, the Sgr~B2 aperture we applied includes
both a position at high-velocity associated with an $x_{1}$ orbit where
\citet{ri+10b} measured [$^{12}$C/$^{13}$C]$\sim 56$ and the Sgr~B2
core where they measured $\sim 4$.  These are similar to the ratios we
measure; in the extended Sgr~B2 aperture [$^{12}$C/$^{13}$C]is $\sim
40$ at $V_{LSR} = +100$~km~s$^{-1}$, and in the Sgr~B2 core is $\sim 4$ at
$V_{LSR} = +50$~km~s$^{-1}$.  The high $^{12}$C/$^{13}$C ratio we measure at
the most red-shifted velocities is thus consistent with the CMZ being
fed with gas with [$^{12}$C/$^{13}$C] ratios typical of those found at
larger distances from the Galactic Centre.  A future investigation
will be to examine the isotopologue line ratios at positions across
the CMZ where less nuclear-processed material may be falling into the
Galactic Centre onto the $x_{1}$ and $x_{2}$ orbits, so as to determine whether
the above result is spread throughout the region, or if it is simply a
special example.

It would also be of interest to examine the CO isotopologue (i.e.\ 
$\rm ^{12}C^{16}O$, $\rm ^{13}C^{16}O$ and $\rm ^{12}C^{18}O$) ratios as a 
function of velocity to determine whether this behaviour is seen for these 
lines in $\rm [^{12}C/^{13}C]$\@.  Furthermore, the 
$\rm ^{12}C^{16}O/^{12}C^{18}O$ ratio would provide a measure of whether the 
same behaviour occurs for $\rm [^{16}O/^{18}O]$, as might be expected if 
infall of less nuclear processed material is occurring.  This is because 
the $\rm [^{16}O/^{18}O]$ gradient has also been found to increase 
with distance from the centre of the Galaxy \citep{wi99}, similarly to that 
of the $\rm [^{12}C/^{13}C]$ ratio.  On the other hand, if the 
$\rm [^{12}C/^{13}C]$ variation arises due to another mechanism (e.g.\ 
differential chemical fractionation), then the $\rm [^{16}O/^{18}O]$ might 
not be found to increase with increasing velocity.

\section{Summary}
\label{sec:summ}

We have mapped a $2.5^{\circ} \times 0.5^{\circ}$ region of the centre
of the Galaxy using the Mopra radio telescope in 18 molecular lines
emitting from 85 to 93\,GHz.  This incorporates most of the region known
as the Central Molecular Zone (CMZ).  The molecular maps have $\sim 40$~arcsec 
spatial resolution and 2~km~s$^{-1}$ spectral resolution, with emission
extending to $V_{LSR}\sim \pm 220$~km~s$^{-1}$.  For eight species the
emission is particularly strong and is widespread across the CMZ: HCN,
HNC, HCO$^{+}$, HNCO, N$_{2}$H$^{+}$, SiO, CH$_{3}$CN and HC$_{3}$N. For the
other molecules mapped the emission is generally confined to the
bright dust cores around Sgr~A, Sgr~B2, Sgr~C and G1.3.  This includes
the isotopologues H$^{13}$CN, HN$^{13}$C and H$^{13}$CO$^{+}$, as well
as c-C$_{3}$H$_{2}$, CH$_{3}$CCH, HOCO$^{+}$, SO and C$_{2}$H.

As seen in several other studies of the CMZ, the molecular emission in
all these species, while widespread, is asymmetric about the centre of
the galaxy, with roughly three-quarters coming from positive
longitudes and one-quarter from negative longitudes.  35--45 percent of the
total molecular emission also arises from the vicinity of the four
brightest dust cores. The overall emission morphology is also
remarkably similar between all molecules. Line profiles are both very
wide and complex, and do vary considerably across the CMZ. While
turbulence contributes to the $\sim 30-50$~km~s$^{-1}$ width of the profiles,
there are also clearly several components along many of the sight
lines.  Absorption from colder, foreground gas is also apparent at
three velocities ($-52, -28$ and $-3$~km~s$^{-1}$) for several locations 
and most strongly in HCN, HNC, and HCO$^{+}$ --
attributable to the sight line passing through spiral arms.

To quantify the overall emission morphology, and its variation between
molecules, we conducted a principal component analysis (PCA) of the
integrated emission from 8 brightest species (as above). The first 4
components are significant, with the first component dominant,
contributing 83 percent of the total variance measured. It essentially
provides an average integrated emission map of the CMZ and
demonstrates that these lines are indeed similar in overall
appearance.  The second component contributes 11 percent to the total
variance and highlights some differences in the bright cores of Sgr~A
and Sgr~B2 between the HCN, HNC and HCO$^{+}$ lines and the other
species.  This is attributable to these three lines being optically
thick in these regions (which we later quantify through an optical
depth analysis).  The third and fourth components (3 and 2 percent of the
variance) highlight smaller differences between SiO, HNCO and
CH$_{3}$CN, reflecting the relative importance of shocks and hot
molecular core emission between some locations.

We have selected apertures around the bright dust cores, as well as
for the total region mapped, in order to study line ratio variations
and to calculate optical depths so that column densities and molecule
masses may be determined. Dense cores species like CH$_{3}$CN, HNCO,
HC$_{3}$N and N$_{2}$H$^{+}$ 
are also found to be more variable with
position, compared to HCN, than are HCO$^{+}$ and HNC.

We also compare the line intensity ratios of HCN, HNC and HCO$^{+}$ to
the same ratios measured for their $^{13}$C isotopologues.  This has
application to the interpretation of these line ratios measured in
extra-galactic sources for determining the excitation conditions,
where the spatial structure remains unresolved. Of note is that the
HCO$^{+}$/HCN ratio ($\sim 0.6$) is found to typically be 2--3 times the
ratio of H$^{13}$CO$^{+}$/H$^{13}$CN.  This is as a result of different
optical depth in the two lines in the $^{12}$C ratio and indicates
that without measurement of the isotopologues the diagnostic ratio
obtained may be in error.  However, in all cases for the CMZ the
diagnostic ratios are found to lie well away from the XDR-excitation
region, and within the PDR region -- as expected given our knowledge
of the overall molecular environment in the centre of the Galaxy.

An optical depth analysis using the $^{13}$C to $^{12}$C isotopologue
line ratios finds modest optical depth for HNC and HCN, with 
$\tau~\sim~1$ for HCO$^{+}$.  Correcting the optical depth allows column
densities and molecular masses to be determined for the CMZ.  Around
50 M$_{\odot}$ of [HCN+HNC+HCO$^{+}$] exists with the region, for
instance.

The ratios of the $^{12}$C to $^{13}$C isotopologues are found to
increase with red-shifted velocity in several locations, indicating a
decreasing optical depth in the line wing.  This is particularly
noticeable for HCO$^{+}$.  However at the extreme red-end 
($V_{LSR} > +100$~km~s$^{-1}$) the ratio exceeds 24, the value determined for 
the $\rm^{12}C/^{13}C$ isotope ratio in the centre of the Galaxy.  This result
has also been found by \citet{ri+10b} for other selected locations in
the centre of the Galaxy.  We discuss whether this provides evidence
for infall of less nuclear-processed gas (where the isotope ratio is
higher) onto the $x_{1}$ and $x_{2}$ orbits in the central regions, so refueling
the gas reservoir in the CMZ.

Line luminosities, relative to that of CO, vary from $\sim 10$~percent for HCN 
to $\sim 1$~percent for SiO. The luminosities are also typically 
$0.1 - 10$~percent of the 
corresponding values that have been measured in other galaxies, though this 
comparison is biased by the selection of the external galaxies studied, as 
well as the comparison of the CMZ to regions several kpc in extent in those 
galaxies.  However, the relative line luminosities (to CO) are not dissimilar, 
with the CMZ lying towards the lower end of the range in HCN/CO ratios 
observed.  This might be attributable to a somewhat lower star formation rate, 
related to the availability of dense gas, than in the comparison galaxies.

The full data set, comprising the data cubes for the 20 emission lines,
is publicly available for further analysis and may be obtained by
contacting the authors.  It will also be made available through the
Australia Telescope National Facility archives (www.atnf.csiro.au).  A
further data set, comprising 24 lines measured in the 7\,mm band
between 42--50~GHz with the Mopra telescope, is also being prepared
for publication.  It includes several lines from lower energy levels
of the same molecules as presented here, so allowing an excitation
analysis to be conducted.

\section*{Acknowledgments}
The Mopra radio telescope is part of the Australia Telescope National Facility 
which is funded by the Commonwealth of Australia for operation as a National 
Facility managed by CSIRO. 
The University of New South Wales Digital Filter Bank used for the observations 
with the Mopra Telescope was provided with support from the Australian Research 
Council (ARC).
We also acknowledge ARC support through Discovery Project
DP0879202. PAJ acknowledges partial support 
from Centro de Astrof\'\i sica FONDAP 15010003 and the GEMINI-CONICYT Fund.
We thank the referee Malcolm Walmsley for his helpful comments and suggestions.



\bsp

\label{lastpage}

\end{document}